\documentclass[a4paper,10pt]{article}
\usepackage{amsfonts}
\usepackage{fancyheadings}
\usepackage[dvips]{graphicx}
\usepackage{graphpap}
\usepackage{ifthen}
\usepackage{amsmath}
\usepackage{amssymb}
\usepackage{makeidx}
\usepackage{fancybox}
\usepackage{psboxit}
\usepackage{amscd,latexsym}
\usepackage{euscript}
\include{psfig}
\usepackage{theorem}
\usepackage[enableskew]{youngtab}
\newfam\scrfam
\batchmode\font\tenscr=rsfs10 \errorstopmode
\ifx\tenscr\nullfont
\else   \font\sevenscr=rsfs7 
        \font\fivescr=rsfs5 
        \skewchar\tenscr='177 \skewchar\sevenscr='177 \skewchar\fivescr='177
        \textfont\scrfam=\tenscr \scriptfont\scrfam=\sevenscr
        \scriptscriptfont\scrfam=\fivescr
        \def\scr{\fam\scrfam}
\fi

\newcommand{\eq}[1]{\begin{equation}#1\end{equation}}              
\newcommand{\eqnono}[1]{\begin{displaymath}#1\end{displaymath}}    
\newcommand{\eqs}[1]{\begin{eqnarray}#1\end{eqnarray}}             
\newcommand{\eqsnono}[1]{\begin{eqnarray*}#1\end{eqnarray*}}       
\newcommand{\mx}[2]{\left(\begin{array}{#1}#2\end{array}\right)}   

\newcommand{\arr}[2]{\begin{array}{#1}#2\end{array}}

\newcommand\rctr{\renewcommand{\theenumi}{(\roman{enumi})}}

\newcommand{\rnum}[1]{\rctr\begin{enumerate}#1\end{enumerate}}

\newcommand{\eqal}[2]{\begin{alignat*}{#1}#2\end{alignat*}}
			\def\scC{{\scr C}}		
\def\d{\delta}

		\def\gog{{\mathfrak g}}

\def\k{\kappa}					
					
\def\l{\lambda}		\def\gol{{\mathfrak l}}	\def\scL{{\scr L}}

		\def\goo{{\mathfrak o}}			
					\def\clP{{\cal P}}
					
				\def\dsR{{\mathbb R}}

\def\o{\omega}

\def\O{\Omega}					
\def\L{\Lambda}

\def\vf{\varphi}

\newcommand\esD{{\EuScript D}}

\newcommand\Lie{\scL}

\newcommand\fr{\over}

\newcommand\ri{{}^\sharp\!}
\newcommand\li{{}^\flat\!}
\newcommand\ct{{}^c\!}

\newcommand{\del}{\nabla}

\newcommand{\minus}{\!{\raise1pt\hbox{$\scriptstyle{-}$}}\!}

\newcommand{\fraction}[1]{{\lower2.5pt\hbox{\eightrm 1}\/\raise2.5pt\hbox{\eightrm #1}}}
\newcommand{\Fraction}[2]{{\lower2.5pt\hbox{\eightrm #1}\/\raise2.5pt\hbox{\eightrm #2}}}
\newcommand{\tr}{\hbox{\rm tr}}

\newcommand{\ul}{\underline}

\newcommand{\bl}{\bar}

\theoremstyle{break}
\newtheorem{theorem}{Theorem}[section]
\newtheorem{lemma}[theorem]{Lemma}
\newtheorem{conjecture}[theorem]{Conjecture}
\newtheorem{proposition}[theorem]{Proposition}
\newtheorem{corollary}[theorem]{Corollary}
\newtheorem{definition}[theorem]{Definition}
\newtheorem{example}[theorem]{Example}
\newtheorem{remark}[theorem]{Remark}
\newtheorem{notation}[theorem]{Notation}
\newtheorem{convention}[theorem]{Convention}

\newcommand{\lem}[2]{\begin{lemma}#2\label{#1}\end{lemma}}

\newcommand{\prop}[2]{\begin{proposition}#2\label{#1}\end{proposition}}

\newcommand{\defn}[2]{\begin{definition}#2\label{#1}\end{definition}}

\newcommand{\nota}[2]{\begin{notation}#2\label{#1}\end{notation}}
		
\newcommand{\exam}[2]{\vskip .1cm\noindent\rule{12.2cm}{.5pt}{\begin{example}\vskip -.4cm\begin{list}{}{\leftmargin 1cm\labelsep .7cm}\item {}\item{\small #2}
	\end{list}\end{example}}\label{#1}\vskip -.5cm\noindent\rule{12.2cm}{.5pt}\vspace{.4cm}}
\newcommand{\proof}[1]{{\em \small proof:}\vspace{-19pt}\begin{list}{}{\leftmargin 1.2cm\labelsep .7cm}\item{\small #1}\hfill$\blacksquare$\end{list}}

\newcommand{\rchapter}[1]{\newpage\label{dummy#1}
	\ifthenelse{\isodd{\pageref{dummy#1}}}{}{\blankpage}
	\chapter{#1}}

\newcommand{\rchaptno}[1]{\newpage\label{dummy#1}
	\ifthenelse{\isodd{\pageref{dummy#1}}}{}{\blankpage}
	\chapter*{#1}}

\newcommand{\ws}{{\scr M}}	\newcommand{\ts}{\ul{\scr M}}
		\newcommand{\tg}{\ul{g}}	\newcommand{\ig}{g}		\newcommand{\nlg}{g'}
		\newcommand{\tE}{\ul{E}}			
						
	\newcommand{\tT}{\ul{T}}					
	\newcommand{\tR}{\ul{R}}					
					\newcommand{\id}{d}		\newcommand{\nd}{d'}
			\newcommand{\ico}{\clP}		\newcommand{\nco}{\clP'\!}
	\newcommand{\tdel}{\ul{\nabla}}

\newcommand{\adel}{\tilde{\tdel}}	\newcommand{\rdel}{\Tilde{\Tilde{\tdel}}}
\newcommand{\aT}{\tilde{\tT}}		\newcommand{\rT}{\Tilde{\Tilde{\tT}}}
\newcommand{\aR}{\tilde{\tR}}		\newcommand{\rR}{\Tilde{\Tilde{\tR}}}
\newcommand{\tsC}{\ul{\scC}}		\newcommand{\aC}{\Tilde{\ul{\scC}}}		\newcommand{\rC}{\Tilde{\Tilde{\ul{\scC}}}}
\newcommand{\rQ}{\Tilde{\Tilde{\ul{Q}}}}\newcommand{\rV}{\Tilde{\Tilde{\ul{V}}}}
\newcommand{\I}{I}

	\newcommand{\tm}{{\ul{m}}}

	\newcommand{\ta}{{\ul{a}}}	
	\newcommand{\tb}{{\ul{b}}}	
\newcommand{\tC}{{\ul{C}}}	\newcommand{\tc}{{\ul{c}}}	
	\newcommand{\td}{{\ul{d}}}	

	\newcommand{\ba}{{\bar{a}}}	
	\newcommand{\bb}{{\bar{b}}}	
	\newcommand{\bc}{{\bar{c}}}	
	\newcommand{\bd}{{\bar{d}}}	

\usepackage[all]{xy}

\begin{document}
%
%
\begin{titlepage}
\title{{\Huge \bf Curvature relations in almost product manifolds}}
\author{
        Magnus Holm\thanks{holm@fy.chalmers.se} $\;$ and Niclas Sandstr\"{o}m\thanks{tfens@fy.chalmers.se}
}
\date{  Institute for Theoretical Physics\\
        G\"oteborg University and Chalmers University of Technology\\
        S-412 96 G\"oteborg, Sweden\\
          \leavevmode
}
\maketitle
%
%
\begin{abstract}
New relations involving curvature components for the various connections appearing in the 
theory of almost product manifolds are given and the conformal behaviour of these 
connections are studied. New identities for the irreducible parts of the deformation 
tensor are derived. Some direct physical applications in Kaluza--Klein and gauge theory are discussed.
\end{abstract}
\end{titlepage}
%
%
\section{Introduction}
In modern day theoretical physics one often deals with additional dimensions besides the ordinary four space-time ones. These extra dimensions manifest
themselves in different forms. In gauge theory they appear as the dimension of the gauge group, in M-theory and string theory, they are required for 
self-consistency. In Kaluza--Klein theory, the gauge theory is obtained by compactification over an internal manifold with an isometry group which equals 
the gauge group. The almost product structure concept makes possible a geometrical formulation which completely describe these theories,
without performing the dimensional reduction. This leads to new insights in their geometrical properties which is unobtainable in the dimensionally
reduced theories themselves. For instance in ref. \cite{Holm98} it was shown that the Nijenhuis tensor of a certain almost product structure measures the 
fieldstrength which in the geometrical language is a measure of the non-integrability of the base manifold of the principal bundle. In almost product 
manifolds, three different connections appear naturally. As is known the Gauss--Codazzi relations connect curvature components of these connections.
In this paper a classification of the relations of all curvature components is given which yields a number of new identities. As a result it becomes 
manifest that the Vidal connection in a principal bundle, or Kaluza--Klein theory, reduces to the gauge-covariant derivative. Since a lot of recent work
\cite{CvYo96,CvDu99} has been made concerning rotating branes which are solutions to various supergravity theories, it will here be stressed that the almost product
manifold vievpoint would be the most geometrical approach to these problems. Direct relations 
for the Ricci tensors in terms of the
characteristic deformation tensors of an almost product structure will be given. These will then be the most natural starting point when making ans\"atze for new solutions
in the supergravity theories.

In the generic case, the Vidal connection will not be metric neither torsion free, and in section 2 we give a review of the theory of general connections. We refer  to \cite{Schouten} for a more detailed treatise in this respect. In section 2 the properties of an arbitrary connection under conformal transformations are also reviewed. Section 3 gives a quick introduction to the basic connections and tensors involved with almost product manifolds. The naturally occuring connections, besides the Levi--Civita one, is the Vidal and adapted connections. All tensors formed from these connections are investigated in section 4.
In that section several new identities are derived, some of which follows directly from the work in ref. \cite{Gr67}, and the conformal properties are studied. In section 5 this is brought to full fruition when the Vidal connection is shown to be identical to the  gauge-covariant derivative in gauge or Kaluza--Klein theory. Possible further developments in this area is discussed in section 6.
%
%
\section{A review on general connections}
This section consists of two parts, the first of which treats a general non-metric connection and its curvature relations together with the 
Bianchi identities. The second part deals with the induced transformation of an arbitrary connection under conformal transformations. 
%
%
\subsection{Non-metric connections}

The most frequently used non-Levi-Civita connections are the ones in which the torsion content is non-zero. In the case of a Vidal connection,
the connection will not in general be metric nor symmetric. Also in the case of embeddings one might encounter non-metric connections while
studying cases with an auxilliary metric on the world-volume. In this subsection a thorough description of connections in the most general case is given.
See also ref. \cite{Schouten}. To this end, the following two important tensor are defined as,
\defn{qandt}{Let $\del$ be a connection in a manifold $\ws$ with non-degenerate metric $\ig$. Now define the torsion tensor, $T$, and the non-metricity
	tensor, $Q$, respectively with characteristics,
	\eqal{1}{
		T:&\quad \L^1 \times \L^1 \longmapsto \L^1\\
		Q:&\quad \L^1 \times \L^1 \times \L^1 \longmapsto \dsR
	}
	by the following equations
	\eqal{1}{
		T(X,Y)&:=\del_XY-\del_YX-[X,Y]\\
		Q(X,Y,Z)&:=(\del_X\ig)(Y,Z)
	}
	where $X,Y,Z\in\L^1$ are vectorfields on $\ws$.
}
A general connection on a manifold with non-degenerate metric can be decomposed into the Levi-Civita connection and
an arbitrary $(2,1)$-tensor. The dimension of this tensor is therefore $m^3$ where $m$ is the dimension of the manifold $\ws$. Below it is shown 
that it can be decomposed into one part containing only the torsion tensor, $T$, and one part containing only the non-metricity tensor, $Q$.
These two tensors have the dimensions ${1\fr 2}m^2(m-1)$ and ${1\fr 2}m^2(m+1)$ respectively which together give $m^3$. The
torsion do not appear directly in the connection but as the contorsion and that is also the case with the non-metricity tensor. The following
notation will be used in what follows,
\eqnono{
	\li T(X,Y,Z):=\ig(T(X,Y),Z)
}
In the next proposition the contorsion and con-metricity tensors are defined.
\defn{contandcoq}{Let $\del$ be a connection in a manifold $\ws$ with non-degenerate metric $\ig$. Define the contorsion tensor, $S$, and the con-metricity
	tensor, $P$, respectively, with same characteristics,
	\eqal{1}{
		S,P:&\quad \L^1 \times \L^1 \times \L^1 \longmapsto \dsR
	}
	by following equations
	\eqal{1}{
		\li S(X,Y,Z)&:={1\fr 2}(\li T(X,Y,Z)-\li T(Y,Z,X)+\li T(Z,X,Y))\\
		\li P(X,Y,Z)&:={1\fr 2}(-Q(X,Y,Z)-Q(Y,Z,X)+Q(Z,X,Y))
	}
	where $X,Y,Z\in\L^1$ are vectorfields on $\ws$ and $S(X,Y)=g^{\minus 1}(\li S(X,Y,\cdot)), P(X,Y)=g^{\minus 1}(\li P(X,Y,\cdot))$.
}
Now, any connection can be expressed in terms of the Levi Civita connection with respect to a non-degenerate metric, denoted by ${}^g\!\del$,
plus the contorsion and the con-metricity tensors defined above, i.e,
\prop{connection}{Let $\del$ be an arbitrary connection on a manifold $\ws$, let further $\ig$ be a non-degenerate metric on $\ws$ and ${}^g\!\del$
       corresponding Levi-Civita connection. Let $S,P$ be the tensors defined in \ref{contandcoq} .Then
       \eqnono{
                \del_XY={}^g\!\del_XY+S(X,Y)+P(X,Y)
      }
}
The curvature tensor of an arbitrary connection, is defined by,
\eq{
      R(X,Y)Z:=[\del_X,\del_Y]Z-\del_{[X,Y]}Z,\label{Rcurvature}
}
will no longer take values in the lie algebra $\goo(m)$ as does the curvature tensor of the Levi-Civita connection, but (will
in the generic case take values) in $\gog\gol(m)$. The identities of the curvature tensor will therefor be altered, and its irreducible parts
look in the generic case like
\eq{
	\yng(1,1)\otimes\yng(1)\otimes\yng(1)=\yng(1,1,1,1)\oplus\yng(2,1,1)\oplus\yng(2,2)\oplus\young(\hfil\hfil,:\hfil,:\hfil)\oplus
						\young(\hfil\hfil\hfil,::\hfil).
}
\prop{curvatureidentities}{The four identities of the Riemann curvature tensor of an arbitrary connection are
	\eqal{2}{
		(i)&\quad&	R_{(ab)c}{}^d=&0\\
		(ii)&\quad&	R_{[abc]}{}^d=&\del_{[a}T_{bc]}{}^d-T_{[ab}{}^eT_{c]e}{}^d\\
		(iii)&\quad&	R_{ab(cd)}=&-(\del_{[a}Q)_{b]cd}-{1\fr 2}T_{ab}{}^eQ_{ecd}\\
		(iv)&\quad&	R_{abcd}-R_{cdab}=&{3\fr 2}(R_{[abc]d}+R_{[bcd]a}-R_{[cda]b}-R_{[dab]c})+\\
					&&&+R_{ab(cd)}-R_{bc(da)}-R_{cd(ab)}+R_{da(bc)}+R_{ac(db)}-R_{db(ac)}
	}
}
By the skew-tableaux (the two tableaux on the right in equation (2) above) it is stressed that these two irreducible parts will vanish when the connection is metric, i.e, the right hand side of identity
3 vanishes. In the generic case there are two possible contractions that can be made.
\defn{ricci}{Let $\del$ be an arbitrary connection on a manifold $\ws$ with curvature tensor, $R$. From the curvature it is possible to construct two types of $(2,0)$ tensors by contraction, namely,
	\eqal{1}{
		R_{ab}:=R_{acb}{}^c,\\
		V_{ab}:=R_{abc}{}^c.
	}
	The first one is the generalized Ricci tensor and the second one will here be refered to as the Schouten two-form.
}
The identities of the Ricci and Schouten tensors can be read off directly from the original curvature identities.
\prop{Ricci-Schouten}{Let $R_{ab}$ be the Ricci tensor and $V_{ab}$ the Schouten tensor of an arbitrary connection, $\del$, then 
	the second and third curvature identities implythe relations,
	\eqal{1}{
		2R_{[ab]}&=V_{ab}-\del_cT_{ab}{}^c-2\del_{[a}T_{b]}-T_{ab}{}^cT_c,\\
		V_{ab}&=-(\del_{[a}Q)_{b]}-{1\fr 2}T_{ab}{}^cQ_c
	}
}
The only integrability conditions to the curvature identities are the Bianchi identities.
\prop{Bianchi}{Let $\del$ be an arbitrary connection with curvature tensor $R$, and torsion tensor $T$. Then the Bianchi identity reads
	\eqnono{
		\del_{[a}R_{bc]d}{}^e=T_{[ab}{}^fR_{c]fd}{}^e
	}
	from which the identities involving the Ricci tensor $R_{ab}$, and the Schouten two-form $V$, are derived,
	\eqal{1}{
		2\del_{[a}R_{b]c}+T_{ab}{}^dR_{dc}&=\del_dR_{abc}{}^d-2T_{d[a}{}^eR_{b]ec}{}^d\\
		dV&=0
	}
}
%
%
\subsection{Conformal transformations}

Below, conformal transformations in the case of an arbitrary connection are studied. There will be some changes compared to the ordinary Levi Civita
case when the connection involves torsion and non-metricity.
\defn{Conformal tensor}{Let $\ws$ be a manifold with metric $\ig$, let further $\del$ be an arbitrary connection on $\ws$. Let $\ct \ig:=e^{2\phi}\ig$
	denote a conformal transformation then define the
	{\bf conformal tensor}, denoted by $\scC$, with characteristics
	\eqal{1}{
		\scC:&\qquad \L^1 \times \L^1 \longmapsto \L^1,
	}
	by
	\eqal{1}{
		\scC(X,Y):=\ct \del_X Y-\del_X Y
	}
	where $X,Y\in\L^1$ are vectorfields on $\ws$.
}
By a straight forward calculation one ends up with the transformations of the characteristic tensors of a connection under conformal transformation.
\prop{conformal transformations}{Let $\ws$ be a manifold with metric $\ig$ and $\del$ be an arbitrary connection on $\ws$. Let $R,T,Q$ denote 
	the Riemann, torsion and non-metricity tensors respectively. Then their transformations under a conformal transformation can be expressed
	in terms of the conformal tensor, $\scC$, as
	\eqal{1}{
		\ct R(X,Y)Z-R(X,Y)Z=&(\del_X\scC)(Y,Z)-(\del_Y\scC)(X,Z)+\\
					&+\scC(X,\scC(Y,Z))-\scC(Y,\scC(X,Z))+\scC(T(X,Y),Z),\\
		\ct T(X,Y)-T(X,Y)=&\scC(X,Y)-\scC(Y,X),\\
		\ct Q(X,Y,Z)-e^{2\phi}Q(X,Y,Z)=&e^{2\phi}[2X[\phi]\ig(Y,Z)-\ig(\scC(X,Y),Z)-\ig(Y,\scC(X,Z))]
	}
}
In the case of the Levi-Civita connection the conformal tensor is most easily extracted from the above proposition.
\prop{Levi-conf}{Let ${}^g\!\del$ be the Levi-Civita connection on a manifold with non-degenerate metric $g$, then its conformal tensor, denoted
       by ${}^g\!\scC$, reads
       \eqnono{
                {}^g\!\scC(X,Y)=X[\phi]Y+Y[\phi]X-g(X,Y)\ri d\phi
       }
}
From these two propositions the conformal tensor in the generic case can be derived.
\prop{generic-conf}{Let $\del$ be an arbitrary connection on a manifold $\ws$ with non-degenerate metric $g$, let further $S,P$ denote
       the contorsion and the con-metricity tensor respectively. Then the conformal tensor of $\del$ reads
       \eqnono{
               \scC(X,Y)={}^g\!\scC(X,Y)+\ct S(X,Y)-S(X,Y)+\ct P(X,Y)-P(X,Y)
       }
}
%
%
\section{The connections associated with an almost product structure}

Here a quick review on the concepts of almost product structures will be given, for a more thorough treatise see refs. \cite{Holm98,Nav84}. 
\nota{nynot}{We will denote the objects on our space with an underline, {\it i.e.},
	\eqal{2}{
		&\ts&	&\text{Manifold}\\
		&T\ts&	&\text{Tangent bundle of $\ts$}\\
		&T^*\ts\qquad&&\text{Cotangent bundle of $\ts$}\\
		&\tg&	&\text{Metric on $\ts$}\\
		&\td&	&\text{Exterior derivative}\\
		&\ul{X}&&\text{Vector field on $\ts$}
	}
	to list the primarily used objects. We will use this underlining principle for all objects on $\ts$ whenever there may be risk of confusion.
}
\defn{i2}{Let I be an almost product structure on a manifold $\ts$ with riemannian metric $\tg$ and let $X,\;Y\in T\ts$ be vector fields. Then the
	triplet ($\ts$, $\tg$, $I$) is called an {\bf riemannian almost product structure} or simply an {\bf almost product manifold} if
	\eqnono{
		\tg(IX,IY)=\tg(X,Y)
	}
	or in other words, $I$ is a automorphism of $\tg$ in the sense that the following diagram commutes:
	\eqnono{
		\xy
			\xymatrix{T\ts\ar[d]_{\tg}\ar[r]^I&T\ts\ar[d]^{\tg}\\
				T^*\ts\ar[r]_{I^t}&T^*\ts}
		\endxy
	}
	{\it i.e.},
	\eqnono{
		I^t\circ\tg\circ\I=\tg
	}
}
\prop{i1}{Let the triplet $(\ts,\tg,I)$ define a riemannian almost product structure on $\ts$ with $\dim \ts=\tm$, then 
	\rnum{
		\item $I^2=1$
		\item All eigenvalues are $\pm1$.
		\item $\tr I=2k-\tm$, where $k$ is the number of positive eigenvalues.
		\item $I\in Gr(k,\tm)\equiv O(\tm)/(O(k)\times O(\tm-k))$.
		\item There is a preferred base called the oriented base in which $I$ is diagonal and ordered, {\it i.e.}, it takes the form
			\eqnono{
				I=\mx{cccccc}{1&&&&&\\
						&\ddots&&&&\\
						&&1&&&\\
						&&&\minus 1&&\\
						&&&&\ddots&\\
						&&&&&\minus 1\\}
			}
	}
}
The almost product structure will serve as a rigging of the tangentbundle by looking at the spaces of eigenvectors to the almost product structure.
\defn{ideffol}{Let I be an almost product structure on $\ts$, then $I$ defines two natural distributions of $T\ts$, denoted $\esD$ and $\esD'$ respectively, 
	in the following way. Let
	\eqsnono{
		\esD_x&:=&\{X\in T_x\ts:IX=X\},\\
		\esD'_x&:=&\{X\in T_x\ts:IX=\minus X\},\\
	}
	then
	\eqnono{
		\esD:=\bigcup_{x\in\ts}\esD_x,\quad\esD':=\bigcup_{x\in\ts}\esD'_x.
	}
}
Seen as an endomorphism of the tangent bundle two projection operators can be formed from the almost product structure as it squares to one.
These will now be projective mappings from the tangent bundle to these two sub-bundles defined above. 
\defn{projs}{From an almost product structure $I$ on a manifold $\ts$ we can define two projection operators through
	\eqsnono{
		\ico&:=&{1\fr 2}(1+I)\\
		\nco&:=&{1\fr 2}(1-I).\\
	}
	These will be mappings in the sense $\ico:T\ts\rightarrow \esD$ and $\nco:T\ts\rightarrow \esD'$ respectively.
}
The Riemann metric in the triplet of a almost product manifold will now split into two parts which will be the induced metrics on these two sub-bundles of the
tangent bundle.
\defn{gandg}{Let $\ts$ be a riemannian or pseudo-riemannian manifold with metric, $\tg$, $I$ a reflective structure with $\ico$ and $\nco$ the
	corresponding projectors, then define the two associated metrics with respect to the reflective structure by
	\eqnono{
		\ig(X,Y):=\tg(\ico X,\ico Y),\quad \nlg(X,Y):=\tg(\nco X,\nco Y)
	}
	which implies that $\tg$ splits into these two parts, {\it i.e.},
	\eqnono{
		\tg=\ig+\nlg.
	}
}
%
%
\subsection{Tensors associated with an almost product structure}

There is one main tensor in the context of almost product manifolds and that is the deformation tensor. This tensor is most suitably decomposed
into two irreducible parts namely the Nijenhuis tensor and the Jordan tensor. 
\defn{nijen2}{Let the triplet $(\ts,\tg,I)$ define a riemannian almost product structure 
	and define the {\bf Nijenhuis tensor} as the measure of how much $\td_I$ fails to be a coboundary
	operator. The Nijenhuis tensor is thus a (2,1) tensor. Let $X,Y\in \L^1$ be vector fields on $\ts$, then the characteristics of the Nijenhuis
	tensor are
	\eqnono{
		N_I(X,Y): \L^1 \times \L^1 \longmapsto \L^1
	}
	and we define it through the quadratic action of $d_I$ on functions $f \in C^\infty (\ts)$,
	\eqnono{
		<-N_I(X,Y), d f>:=\td_\I \td_\I f(X,Y).
	}
	It follows that the Nijenhuis tensor measures the failure in closure of the operator $d_I$ and can thus be 
	considered as a kind of torsion. Alternatively, as the equivalent definition below shows, it measures the
	curvature of the endomorphism, {\it i.e.},
	\eqnono{
		N_\I[X,Y]:=I([X,Y]_\I)-[I(X),I(Y)]
	}
	Alternatively the Nijenhuis tensor can be seen as measuring how far this endomorphism is from being a Lie algebra homomorphism
	of the infinite-dimensional Lie algebra of vector fields on $\ws$.
}
\defn{jordantensor}{Let the triplet $(\ts,\tg,I)$ define a riemannian almost product structure, and let $\{\cdot,\cdot\}$ be the Jordan bracket.
	The {\bf Jordan tensor} associated to $I$, denoted $M_I$, with the following characteristics:
	\eqnono{
		M_I: \L^1\times\L^1 \longmapsto \L^1
	}
	is defined by,
	\eqnono{
		M_I(X,Y):=I\{X,Y\}_I-\{IX,IY\}
	}
	where $X,Y\in\L^1$ are vector fields on $\ts$. The analogy to the Nijenhuis tensor is obvious .
}
Both the Nijenhuis and the Jordan tenor can be expressed entirely in terms of the covariant derivative of the almost product structure.
\defn{Ideformation}{Let the triplet $(\ts,\tg,I)$ define a riemannian almost product structure. Let $\tdel$ be the Levi--Civita connection on $\ts$ and 
	define the {\bf deformation tensor} associated with
	the endomorphism $I$, denoted $H_I$, with the following characteristics:
	\eqnono{
		H_I:\L^1 \times \L^1 \longmapsto \L^1
	}
	$H_I$ is defined by the expression
	\eqnono{
		H_I(X,Y):=(I\tdel_X I-\tdel_{IX}I)(Y),
	}
	where $X,Y\in \L^1$ are two vector fields on $\ts$. An equivalent definition is given by,
	\eqnono{
		H_I(X,Y):=N_I(X,Y)+M_I(X,Y).
	}
}
Looking at the characteristic tensors of a distribution the deformation tensor of an almost product structure can now be decomposed into the 
deformation tensors of the two complementary distributions defined by the almost product structure.
\defn{deform}{Let $\esD$ be a $k$-distribution with projection $\ico$ on a riemannian manifold $\ts$ with non-degenerate metric $\tg$. Let $\tdel$ be
        the Levi--Civita connection with respect to this metric and let $\nco:=1-\ico$ be the coprojection of $\esD$. 
        Now define the following tensors with characteristics
        \eqal{2}{
                H,L,K:&\quad&\L^1_\esD\times\L^1_\esD &\longmapsto \L^1_{\esD'}\\
                \k:&&\L^1_{\esD'}&\longmapsto \dsR
        }
        and
        \eqal{3}{
                (i)&&   \quad H(X,Y)    &:=\nco\tdel_{\ico X}\ico Y      &\quad&\text{\bf deformation tensor},\\
                (ii)&&  L(X,Y)       &:=\frac{1}{2}(H(X,Y)-H(Y,X))       &&\text{\bf twisting tensor},\\
                (iii)&& K(X,Y)       &:=\frac{1}{2}(H(X,Y)+H(Y,X))      &&\text{\bf extrinsic curvature tensor},\\
                (iv)&&  \ri\k   &:=\tr H                                &&\text{\bf mean curvature tensor},\\
                (v)&&   W(X,Y)       &:=K(X,Y)-{1\fr k}\ri\k \ig(X,Y)   &&\text{\bf conformation tensor}.
        }
        This gives us the decomposition of the deformation tensor in its anti-symmetric, symmetric-traceless and trace parts accordingly,
        \eqnono{
                H=L+W+{1\fr k}\ri\k\ig.
        }
}
The extrinsic curvature tensor and the twisting tensor can be written in a more elegant fashion.
\prop{defrel}{Let $\esD$ be a distribution on a manifold $\ts$ with metric $\tg$, let further $g(X,Y)=\tg(\ico X,\ico Y)$ be the induced metric on the
        distribution, then the symmetric part of the deformation tensor can be written like
        \eqnono{
                K(X,Y)(\vf)=\minus{1\fr 2}\Lie_{\ri\vf'}\ig(X,Y),\;or\quad\li K(X,Y,Z)=\minus{1\fr 2}\Lie_{Z'}\ig(X,Y),
        }
        where the prime denotes projection along the normal directions by $\nco$. The relation for the anti-symmetric part on the other hand is
        \eqnono{
                L(X,Y)={1\fr 2}\nco [\ico X,\ico Y]
        } 
}
The conformal properties of the irreducible parts of the deformation tensor can be found in next proposition.
\prop{defrela}{Let $\ts$ be a riemannian manifold with metric $\tg$, let $I$ be an almost product structure on $\ts$ which split the metric in 
	$\tg=\ig+\nlg$ and let $\l=e^{2\phi}$ be a conformal transformation on $\ig$, {\it i.e.}, $\ct\tg=\l\tg$ then the symmetric parts of the 
        deformation tensor will transform like
        \eqsnono{
                \ct K(\vf)&=&K(\vf)-\frac{1}{2}\l^{\minus 1}\ri\vf'[\l]g=K(\vf)-\ri\vf[\phi]g\\
                \ct\k(X)&=&\k(X)-\frac{1}{2}k\l^{\minus 1} X'[\l]=\k(X)-kX'[\phi]\\
                \ct W&=&W\\
                \ct L&=&L
        }
}
Now denoting the deformation tensor of the complementary distribution $\esD'$ by $H'$ and its irreducible parts by $L',K',\k',W'$ respectively
we can express the Nijenhuis tensor and the Jordan tensor in terms of these characteristic tensors.
\lem{nijenigen}{Let $I$ be an almost product structure on a manifold $\ts$ and let its associated projection operators be
	$\ico:={1\fr 2}(1+I)$, $\nco:={1\fr 2}(1-I)$, then
	\eqal{2}{
		(i)&&\qquad	N_{\ico}=&N_{\nco}\cr
		(ii)&&		N_I=&4N_{\ico}\cr
		(iii)&&		{1\fr 2}[\ico,\nco ]=&N_{\ico}\cr
		(iv)&&\qquad	N_{\ico}(X,Y)=&-\nco [\ico X,\ico Y]-\ico [\nco X, \nco Y]
	}
}
\prop{nijol}{Let the triplet $(\ts,\tg,I)$ define an riemannian almost product structure and let $L,\; L'$ be the twisting tensors of the distributions defined by $I$.
	Then
	\eqnono{
		{1\fr 8}N_I=-L-L'
	}
}
\lem{almostjordan}{Let $I$ be an almost product structure on a manifold $\ts$ and let its associated projection operators be
	$\ico:={1\fr 2}(1+I)$, $\nco:={1\fr 2}(1-I)$. Let $M$ denote the Jordan tensor, then
	\eqal{2}{
		(i)&&\qquad	M_{\ico}=&M_{\nco}\cr
		(ii)&&		M_I=&4M_{\ico}\cr
		(iii)&&		{1\fr 2}\{\ico,\nco \}=&M_{\ico}\cr
		(iv)&&\qquad	M_{\ico}(X,Y)=&-\nco \{\ico X,\ico Y\}-\ico \{\nco X, \nco Y\}
	}
}
\prop{almostjordan2}{Let the triplet $(\ts,\tg,I)$ define a riemannian almost product structure, $K,\; K'$ be the extrinsic curvature tensors 
of the distributions defined by $I$,
	then
	\eqnono{
		{1\fr 8}M_I=-K-K'
	}
}
%
%
\subsection{Three relevant connections}

In a almost product manifold there are three different connections of importance, which will be defined in this subsection. The first is of course the
Levi--Civita connection, from which the other two will be defined by simply adding a tensor to it. These will be refered to as the adapted and the Vidal
connection. Their basic feature is that they commute with the almost product structure which means that they respect the rigging of the tangent space
defined by the almost product structure. The additional feature of the adapted connection is that it is metric which together with the above feature 
implies that it respects the induced metrics on the two characteristic distributions associated with the almost product structure. The Vidal connection
which is metric iff the characteristic distributions are geodesic will play an important role when looking at gauge theories and fiber bundles since
they need no metric in the total space and are of the type $(GF,GD)$. By adding the group metric to the fiber we can construct an almost product manifold
in which the Vidal connection will reduce to the gauge covariant derivative. This will be explicitly done in section 5. The curvature components of
the Vidal connections lying entirely in the tensor algebra of the characteristic distributions, also called the semi-basic parts, will in a more
natural way measure the curvature in the respective distributions. This is due to the fact that it does not depend on the connections in its co-parts. What this means
explicitly will become clear when the relations are derived. 
\defn{adap}{Let $\ts$ be a riemannian or pseudo-riemannian manifold with non-degenerate metric $\tg$ and corresponding Levi--Civita connection $\tdel$.
	Let $\I$ define distributions as in definition \ref{ideffol}. Then the following two definitions of the {\bf adapted connection} are equivalent
	\rnum{
	\item	$\adel_XY:=\tdel_XY+A(X,Y), \quad A(X,Y):={1\fr 2}\I\tdel_X\I(Y)$
	\item	$\adel_XY:=\ico\tdel_X\ico Y+\nco\tdel_X\nco Y$
	}
}
\defn{rdel}{Let $\ts$ be a riemannian or pseudo-riemannian manifold with non-degenerate metric $\tg$ and corresponding Levi--Civita connection $\tdel$,
	let $\I$ define a foliation. Then the {\bf Vidal connection} is defined by
	\eqnono{
		\rdel_XY:=\adel_XY+B(X,Y),\quad B(X,Y):={1\fr 4}(\tdel_{IY}I+I\tdel_YI)(X).
	}
}
\prop{Vidalconnection}{Let $\ul{X},\ul{Y}$ be vectorfields on an almost product manifold $(\ts,\tg,I)$, let $X=\ico \ul{X},X'=\nco\ul{X}$ and
	similar for $\ul{Y}$. Then the Vidal connection can be written
	\eqnono{
		\rdel_{\ul{X}}\ul{Y}=\mx{cc}{\ico \tdel_X Y& \nco [X,Y']\\\ico [X',Y]&\nco\tdel_{X'}Y'}
	}
}
The two recently introduced tensors, $A$ and $B$, are in fact related.
\prop{boa}{Let $B$ be the tensor defined in \ref{rdel} and $A$ the tensor defined in \ref{adap}, then it is possible express the tensor $B$ in terms 
 of $A$ and the almost product structure $I$ as
	\eqnono{
		B(X,Y)={1\fr 2}\left(A(Y,X)-A(IY,IX)\right).
	}
}
The most important property of the two connections defined above, is that they both commute with the almost product structure.
\prop{aroni}{Let $\adel$ denote the adapted connection defined in \ref{adap} and $\rdel$ the Vidal connection defined in \ref{rdel}, then
	their principal feature is that they both commute with the almost product structure $I$, {\it i.e.},
	\eqnono{
		\adel_XI=\rdel_XI=0
	}
}
Only one of them though will, in the generic case, be metric and that is the adapted connection.
\prop{adelmetric}{Let the triplet $(\ts,\tg,I)$ be a riemannian almost product structure on $\ts$ and $\adel$ the adapted connection
	defined in \ref{adap}, then this connection is metric with respect to the splitting of $\tg$ according to \ref{gandg}, {\it i.e.},
	\eqal{1}{
		\adel \ig&=0\\
		\adel \nlg&=0
	}
}
The connection components takes a most pleasant form in the oriented basis. The notation, $$\tE_{\ba}=(\tE_a,\tE_{a'})$$
will be used,
where unprimed(primed) index denotes the basis of the characteristic unprimed(primed) distribution.
\prop{del}{Let the triplet $(\ts,\tg,I)$ define a riemannian almost product structure, let $\ul{\o}$, $\tilde{\ul{\o}}$ and $\Tilde{\Tilde{\ul{\o}}}$
	denote the connection one-forms of the Levi--Civita connection, the adapted connection and the Vidal connection respectively.
	Let furthermore $H,H'$ denote the deformation tensors with respect to $I$ and $C,C'$ be coefficients of anholonomy Then
	\eqsnono{
		\ul{\o}&=&\left[\mx{cc}{\o&H\\-H^t&\O},\mx{cc}{\O'&H'\\-{H'}^t&\o'}\right]\\
		\tilde{\ul{\o}}&=&\left[\mx{cc}{\o&\;\;0\;\;\\\;\;0\;\;&\O},\mx{cc}{\O'&\;\;0\;\;\\\;\;0\;\;&\o'}\right]\\
		\tilde{\tilde{\ul{\o}}}&=&\left[\mx{cc}{\o&\;\;0\;\;\\\;\;0\;\;&C},\mx{cc}{C'&\;\;0\;\;\\\;\;0\;\;&\o'}\right]\\
	}
}
The coefficients of anholonomy are defined by $[E_{\ba},E_{\bb}]=:C_{\ba\bb}{}^{\bc}E_{\bc}$ and can be used to express all the connection
components and the deformation tensors. It is interesting to note their behaviour under a local $O(k)\times O(k')$ transformation:
\eqal{2}{
	\tilde{C}_{ab}{}^{c}=&u_a{}^du_b{}^eC_{de}{}^fu^{\minus 1}_f{}^c+2\tilde{E}_{[a}[u_{b]}{}^f]u^{\minus 1}_f{}^c,&\quad&
				\text{Coefficients of anholonomy}\\
	\tilde{C}_{ab}{}^{c'}=&u_a{}^du_b{}^eC_{de}{}^{f'}u^{\minus 1}_{f'}{}^{c'},&\quad&\text{Tensor}\\
	\tilde{C}_{a'b}{}^{c}=&u_{a'}{}^{d'}u_b{}^eC_{d'e}{}^fu^{\minus 1}_f{}^c+\tilde{E}_{a'}[u_{b}{}^f]u^{\minus 1}_f{}^c,&\quad&\text{Connection}\\
}
This is what enables the defininition of the Vidal connection, i.e. the observation that the $C_{ab'}{}^{c'},C_{a'b}{}^{c}$ parts
of the structure coefficients transform as connections under local $O(k)\times O(k')$ transformations.
\prop{sructure-connection}{Let $\o,\O,H,L,K$ be the components of the Levi-Civita connection and $C$ the coefficients of anholonomy, then
	they are related through the following relations,
	\eqal{1}{
		C_{abc}&=2\o_{[ab]c}\\
		C_{a'bc}&=\O'_{a'bc}+H_{bca'}\\
		C_{abc'}&=2L_{abc'}\\
		C_{a'b'c}&=2L'_{a'b'c}\\
		C_{ab'c'}&=\O_{ab'c'}+H'_{b'c'a}\\
		C_{a'b'c'}&=2\o'_{[a'b']c'}
	}
	with inverse relations
	\eqal{1}{
		\o_{abc}&=C_{a[bc]}-{1\fr 2}C_{bca}={1\fr 2}C_{abc}+C_{c(ab)}\\
		\O'_{a'bc}&=C_{a'[bc]}-{1\fr 2}C_{bca'}\\
		K_{abc'}&=C_{c'(ab)}\\
		L_{abc'}&={1\fr 2}C_{abc'}\\
		H_{abc'}&={1\fr 2}C_{abc'}+C_{c'(ab)}\\
		H'_{a'b'c}&={1\fr 2}C_{a'b'c}+C_{c(a'b')}\\
		L'_{a'b'c}&={1\fr 2}C_{a'b'c}\\
		K'_{a'b'c}&=C_{c(a'b')}\\
		\O_{ab'c'}&=C_{a[b'c']}-{1\fr 2}C_{b'c'a}\\
		\o'_{a'b'c'}&=C_{a'[b'c']}-{1\fr 2}C_{b'c'a'}={1\fr 2}C_{a'b'c'}+C_{c'(a'b')}\\
	}
		
}
%
%
\section{The curvature components and their relations}

As fiber bundles and fibrations are examples of almost product manifolds with the additional property of the existense of a surjective submersion
of the total space down to a base space, it would be interesting to see what parts of the Vidal and adapted connections, which are defined on the
total space, that survive under this submersion. In the total space two new differential operators were defined in $\cite{Holm98}$.
\defn{nybound}{Let $I$ be an almost product structure on a manifold $\ts$ with exterior derivative $\td$. Let furthermore $\td_I$ denote
	the exterior derivative associated with $I$ and define two new differential operators by
	\eqsnono{
		\id&:=&{1\fr 2}(\td+\td_\I)\\
		\nd&:=&{1\fr 2}(\td-\td_\I)
	}
	An equivalent definition is by the two projection operators defined by the endomorphism $\I$, $\ico:={1\fr 2}(1+I)$ and
	$\nco:={1\fr 2}(1-I)$, then the operators are simply $\id\equiv\td_\ico$ and $\nd\equiv\td_{\nco}$.
}
These differential operators become coboundary operators if and only if the Nijenhuis tensor vanishes, which is the same as to say that both
the characteristic distributions of the almost product structure are integrable. In a fibration for instance this is not normally true, except
for the trivial case of a product manifold, so we will keep track of all components surviving the submersion and those who will not. When it comes
to these differential operators it is therefor clear that $d$ defined above will in general differ from the exterior derivative defined on the base space.
By projecting out the semi-basic parts of all quantities we can keep track of the parts that survives the submersion. 
\defn{distributionbrackets}{Let the triplet $(\ts,\tg,I)$ denote a riemannian almost product structure and $\esD$, $\esD'$ be the associated
	distributions then define the brackets associated with these distributions with following characteristics
	\eqal{1}{
		[\;\cdot\;,\;\cdot\;]^{\esD}:&\quad \L_{\esD}^1\times \L_{\esD}^1 \longmapsto \L^1_{\esD},\\
		[\;\cdot\;,\;\cdot\;]^{\esD'}:&\quad \L_{\esD'}^1\times \L_{\esD'}^1 \longmapsto \L^1_{\esD'},
	}
	by
	\eqal{1}{
		[X,Y]^{\esD}&:=\ico [\ico X, \ico Y],\\
		[X,Y]^{\esD'}&:=\nco [\nco X, \nco Y],
	}
	where $X,Y \in \L^1$ are vectorfields on $\ts$.
}
Here it is clear that the twisting tensor that measures the amount of non-commutativity is non-semi-basic. The two brackets defined above will therefor
not satisfy the Jacobi identity in the total space, but will differ with some terms involving the twisting tensors. The same procedure can be made
to define the semi-basic torsions and curvature tensor of a distribution.
\defn{vert-horiz}{Let the triplet $(\ts,\tg,I)$ denote a riemannian almost product structure and $\esD$, $\esD'$ be the associated
	distributions. Let further $T^p_q(T\ts)$ denote the set of $(p,q)$-tensors on $\ts$ and $T^p_q(\esD)$ ($T^p_q(\esD')$) denote the set
	of tensors lying entirely in $\esD$ ($\esD'$). Now define the associated Levi--Civita connections with following characteristics
	\eqal{1}{
		\tdel^{\esD}_X:&\quad T^p_q(T\ts)\longmapsto T^p_q(\esD),\\
		\tdel^{\esD'}_X:&\quad T^p_q(T\ts)\longmapsto T^p_q(\esD'),
	}
	through
	\eqal{1}{
		\tdel^{\esD}_X Y:=&\ico \tdel_{\ico X} \ico Y,\\
		\tdel^{\esD'}_X:=&\nco \tdel_{\nco X} \nco Y,
	}
	where $X,Y \in \L^1$ are vectorfields on $\ts$. Further define the torsion and curvature of the corresponding connections by
	\eqal{1}{
		T^{\esD}(X,Y)&:=\tdel^{\esD}_X Y- \tdel^{\esD}_Y X-[X,Y]^{\esD},\\
		T^{\esD'}(X,Y)&:=\tdel^{\esD'}_X Y- \tdel^{\esD'}_Y X-[X,Y]^{\esD'},
	}
	and
	\eqal{1}{
		R^{\esD}(X,Y)Z&:=\tdel^{\esD}_X \tdel^{\esD}_Y Z- \tdel^{\esD}_Y \tdel^{\esD}_X Z-\tdel^{\esD}_{[X,Y]^{\esD}}Z,\\
		R^{\esD'}(X,Y)Z&:=\tdel^{\esD'}_X \tdel^{\esD'}_Y Z- \tdel^{\esD'}_Y \tdel^{\esD'}_X Z-\tdel^{\esD'}_{[X,Y]^{\esD'}}Z.
	}
}
In this case it is clear that the curvature defined above, will not in general be tensorial in the latter indices. As before these non-tensorial parts
will vanish under the submersion. The torsion will still be tensorial though.
%
%
\subsection{The Vidal connection}

In this subsection, all tensors associated with the Vidal connection will be derived, that is the torsion tensor, the non-metricity tensor,
the Riemann tensor and its traces. The curvature identities are used to express all components but the two totally semi-basic ones only in terms
of the different irreducible parts of the deformation tensor and its derivatives. From these curvature identities there also arise a couple of new
relations involving only parts of the deformation tensor. Two of these will evidently become the Bianchi identity of the two twisting tensors but two
others will appear in a more unfamiliar fashion.

From the definition it is clear that the Vidal connection is neither torsion-free, nor metric in the generic case of an almost product
manifold. It is therefor interesting to see what the torsion and non-metricity tensor look like in this case. 
In ref. \cite{Holm98} the following proposition was derived. 
\prop{rtor}{Let the triplet $(\ts,\tg,I)$ define an almost product manifold, let $N_I$ denote the Nijenhuis tensor of $I$ and 
	$\rdel$ denote the Vidal connection defined in \ref{rdel}, then,
	\eqnono{
		{1\fr 4}N_I(X,Y)=\rT(X,Y).
	}
}
Together with proposition \ref{nijol} the torsion tensor can be written in component form.
\prop{vidal torsion}{Let $\rT$ be the torsion tensor of the Vidal connection, $\rdel$, then in component form it reads
	\eqal{1}{
		\rT_{ab}{}^{c}=&0,\\
		\rT_{ab}{}^{c'}=&-2L_{ab}{}^{c'},\\
		\rT_{a'b}{}^{c}=&0,\\
		\rT_{ab'}{}^{c'}=&0,\\
		\rT_{a'b'}{}^{c}=&-2L'_{a'b'}{}^{c},\\
		\rT_{a'b'}{}^{c'}=&0.
	}
	and
	\eqal{1}{
		\rT_a=&0,\\
		\rT_{a'}=&0.
	}
}
As was seen in \cite{Holm98} the torsion tensor measures the non-integrability of the two complementary distributions defined by an almost
product structure. The non-metricity of the Vidal connection is put in the next proposition.
\prop{vidal-metric}{Let the triple $(\ts,\tg,I)$ denote a riemannian almost product structure with associated metrical decomposition $\tg=\ig+\nlg$.
	Let further $\rdel$ denote the Vidal-connection then the following relations hold
	\eqal{1}{
		(\rdel_Z\ig)(X,Y)&=0\\
		(\rdel_{Z'}\ig)(X,Y)&=-2K_{Z'}(X,Y)=(\Lie_{Z'}\ig)(X,Y)\\
		(\rdel_Z\nlg)(X',Y')&=-2K'_{Z}(X',Y')=(\Lie_{Z}\nlg)(X',Y')\\
		(\rdel_{Z'}\nlg)(X',Y')&=0\\
	}
}
In component form the non-metricity tensor can be read off from the next proposition.
\prop{vidal non-metricity}{Let $\rQ$ be the non-metricity tensor of the Vidal connection, $\rdel$, then in component form it reads
	\eqal{1}{
		\rQ_{abc}=&0\\
		\rQ_{a'bc}=&-2K_{bca'}\\
		\rQ_{abc'}=&0\\
		\rQ_{a'b'c}=&0\\
		\rQ_{ab'c'}=&-2K'_{b'c'a}\\
		\rQ_{a'b'c'}=&0
	}
	The two traces following from a three tensor symmetric in two indices is found to be
	\eqal{1}{
		\rQ^1_a=&0\\
		\rQ^1_{a'}=&0\\
		\rQ_a:=&\rQ^2_a=-2\k'_a\\
		\rQ_{a'}:=&\rQ^2_{a'}=-2\k_{a'}\\
	}
}
The basic curvature relations are found from definitions \ref{Rcurvature} and \ref{vert-horiz}, they are given in the proposition below,
\prop{vidalnoindex}{Let $\rdel$ be the Vidal connection and $\rR$ denote the curvature tensor theorof then the different parts reads
	\eqal{1}{
		\rR(X,Y)Z&=R^{\esD}(X,Y)Z-2\ico [L(X,Y),Z]\\
		\rR(X,Y)Z'&=-2(\rdel_{Z'}L)(X,Y)\\
		\rR(X,Y')Z&=\ico\del_{X}\ico[Y',Z]-\ico[Y',\ico\del_{X}Z]-\ico\del_{\ico[X,Y']}Z-\ico[\nco[X,Y'],Z]\\
	}
}
In order to deal with the curvature identities it is convenient to put these relations in component form.
\prop{vidalcurvatures}{Let $\rR$ be the Riemann-tensor with respect to the Vidal-connection, then $\rR$ has the following components,
	\eqal{1}{
		\rR_{abc}{}^d=&R^{\esD}_{abc}{}^d-2L_{ab}{}^{e'} C_{e'c}{}^d\\
                \rR_{abc}{}^{d'}=&0\\
                \rR_{abc'}{}^d=&0\\
                \rR_{abc'}{}^{d'}=&-2(\rdel_{c'}L)_{ab}{}^{d'}\\
                \rR_{a'bc}{}^d=&E_{a'}[{\o}_{bc}{}^d]-E_b[C_{a'c}{}^d]-C_{a'b}{}^e{\o}_{ec}{}^d-C_{a'b}{}^{e'}C_{e'c}{}^{d}-
				C_{a'c}{}^e{\o}_{be}{}^d+\o_{bc}{}^eC_{a'e}{}^d\\
                \rR_{a'bc}{}^{d'}=&0\\
                \rR_{a'bc'}{}^d=&0\\
                \rR_{a'bc'}{}^{d'}=&E_{a'}[{C}_{bc'}{}^{d'}]-E_{b}[\o_{a'c'}{}^{d'}]-C_{a'b}{}^{e'}{\o}_{e'c'}{}^{d'}-C_{a'b}{}^{e}C_{ec'}{}^{d'}-
				\o_{a'c'}{}^{e'}{C}_{be'}{}^{d'}+C_{bc'}{}^{e'}\o_{a'e'}{}^{d'}\\
                \rR_{a'b'c}{}^d=&-2(\rdel_{c}L)_{a'b'}{}^d\\
                \rR_{a'b'c}{}^{d'}=&0\\
                \rR_{a'b'c'}{}^d=&0\\
                \rR_{a'b'c'}{}^{d'}=&R^{\esD'}_{a'b'c'}{}^{d'}-2L_{a'b'}{}^{e}C_{ec'}{}^{d'}
       }
}
This is the ``raw'' expressions for the curvature components of the Vidal connection, but after using the identities of the Riemann tensor seen in section 
\ref{curvatureidentities}
these will simplify remarkably. There will also appear a couple of identities involving only parts of the deformation tensor. Starting with the second identity
it can be seen, using the Tic-Tac-Toe notation introduced in appendix A, that the second identity can be split according to the rigging into 8 irreducible parts, corresponding
to the following young tableaux.
\eqal{1}{
	\yng(1,1,1)\otimes\yng(1)=&\left(\young(o,o,o)\otimes\young(o)\right)\oplus\left(\young(o,o,o)\otimes\young(x)\right)\oplus\left(\young(o,o,x)\otimes\young(o)\right)\oplus
		\left(\young(o,o,x)\otimes\young(x)\right)\oplus\\
		&\left(\young(o,x,x)\otimes\young(o)\right)\oplus\left(\young(o,x,x)\otimes\young(x)\right)\oplus
		\left(\young(x,x,x)\otimes\young(o)\right)\oplus\left(\young(x,x,x)\otimes\young(x)\right)
}
Here will be listed only the first half of the identities as they of course are symmetric upon changing primes and unprimes.
\eqs{
	\rR_{[abc]}{}^{d}&=&0\\
	\rdel_{[a}L_{bc]}{}^{d'}&=&0\\
	\rR_{c'[ab]}{}{d}&=&-L_{ab}{}^{e'}L'_{c'e'}{}^{d}\\
	\rR_{abc'}{}^{d'}&=&-2\rdel_{c'}L_{ab}{}^{d'}
}
So in conclusion the first identity is structurally inherited by the totally semi-basic part of the Vidal curvature, the second leads to a bianchi
identity for the twisting tensor, the third relating the anti-symmetric part of $\rR_{c'[ab]}{}^{d}$ in terms of the twisting tensors and the fourth
a faster way of deriving the $\rR_{abc'}{}^{d'}$ component. 
Thus, the following proposition is proved:
\prop{twistbianchi}{Let $\rdel$ be the Vidal connection associated with a almost product structure and $L,L'$ be the respective twisting tensors of the
       associated distributions then the following Bianchi identities hold,
\eqal{1}{
                \rdel_{[a}L_{bc]}{}^{d'}=&0\\
                \rdel_{[a'}L'_{b'c']}{}^{d}=&0
}
}
The third identity which is decomposed as,
\eqal{1}{
	\yng(1,1)\otimes\yng(2)=&\left(\young(o,o)\otimes\young(oo)\right)\oplus\left(\young(o,x)\otimes\young(oo)\right)\oplus\left(\young(o,o)\otimes\young(ox)\right)\oplus\\
		&
			\left(\young(o,o)\otimes\young(xx)\right)\oplus\left(\young(o,x)\otimes\young(ox)\right)\oplus
		\left(\young(x,x)\otimes\young(oo)\right)\oplus\\&\left(\young(x,x)\otimes\young(ox)\right)\oplus\left(\young(o,x)\otimes\young(xx)\right)\oplus
			\left(\young(x,x)\otimes\young(xx)\right)
}
In this case there are 9 irreducible parts of which five will be listed and the other follows due to symmetry.
\eqs{
	\rR_{ab(cd)}&=&-2L_{ab}{}^{e'}K_{cde'}\\
	\rR_{a'b(cd)}&=&-(\rdel_{b}K)_{cda'}\\
	0&=&0,\\
	\rR_{ab(c'd')}&=&2(\rdel_{[a}K')_{c'd'|b]}\\
	0&=&0
}
The first identity here gives no new information, the second gives yet another part of the $\rR_{a'bcd}$ component, the third and the fifth contain
nothing while the fourth together with the original expression for the $\rR_{abc'd'}$ component gives a new non-trivial identity which proves the next
proposition.
\prop{vidal-identity}{Let $\rdel$ be the Vidal-connection, $K,L,K',L'$ be the second fundamental tensors of a almost product structure then we have
	the following identities
	\eqal{1}{
			\rdel_{[Z}K'_{W]}(X',Y')+\rdel_{(X'}L_{Y')}(Z,W)&=0\\
			\rdel_{[Z'}K_{W']}(X,Y)+\rdel_{(X}L'_{Y)}(Z',W')&=0
	}
	In component form the same expressions read
	\eqal{1}{
		(\rdel_{[a}K')_{c'd'|b]}+(\rdel_{(c'}L)_{ab|d')}=&0\\
		(\rdel_{[a'}K)_{cd|b']}+(\rdel_{(c}L')_{a'b'|d)}=&0
	}
	Contracted the identities read
	\eqal{1}{
		(\rdel_{[a}\k')_{b]}+(\rdel_{c'}L)_{ab}{}^{c'}=&0\\
		(\rdel_{[a'}\k)_{b']}+(\rdel_{c}L)_{a'b'}{}^{c}=&0
	}
}
Here the notation $$L_{X'}(Y,Z):=\tg(L(Y,Z),X')$$ is used.
The last identity of the Riemann curvature is the most non-trivial of them all. It can be decomposed through the box symmetry, i.e.
\eqal{1}{
	\yng(2,2)=\young(oo,oo)\oplus\young(oo,ox)\oplus\young(oo,xx)\oplus\young(ox,ox)\oplus\young(xx,xo)\oplus\young(xx,xx).
}
The only identity of these which gives new information is the second why it is the only one listed. This identity though gives the opportunity
of writning the entire $\rR_{a'bcd}$ component purely in terms of the different parts of the deformation tensor and its derivatives.
\eq{
	\rR_{a'bcd}=-(\rdel_bK)_{cda'}-2(\rdel_{[c}K)_{d]ba'}+2L_{cd}{}^{e'}L'_{a'e'b}-4L_{b[c}{}^{e'}L'_{a'e'|d]}
}
This relation together with the identity in \ref{vidal-identity} prooves the following proposition which is the final form of the Vidal curvature
components.
\prop{vidal curvature}{Let $\rR$ be the curvature tensor of the Vidal connection then its components can be written
	\eqal{1}{
		\rR_{abcd}=&R^{\esD}_{abcd}-2L_{ab}{}^{e'} C_{e'cd}\\
                \rR_{abcd'}=&0\\
                \rR_{abc'd}=&0\\
                \rR_{abc'd'}=&2[(\rdel_{[a}K')_{c'd'|b]}-(\rdel_{[c'}L)_{ab|d']}]\\
                \rR_{a'bcd}=&-(\rdel_bK)_{cda'}-2(\rdel_{[c}K)_{d]ba'}+2L_{cd}{}^{e'}L'_{a'e'b}-4L_{b[c}{}^{e'}L'_{a'e'|d]}\\
                \rR_{a'bcd'}=&0\\
                \rR_{a'bc'd}=&0\\
                \rR_{a'bc'd'}=&(\rdel_{a'}K')_{c'd'b}+2(\rdel_{[c'}K')_{d']a'b}-2L'_{c'd'}{}^{e}L_{bea'}+4L'_{a'[c'}{}^{e}L_{be|d']}\\
                \rR_{a'b'cd}=&2[(\rdel_{[a'}K)_{cd|b']}-(\rdel_{[c}L')_{a'b'|d]}],\\
                \rR_{a'b'cd'}=&0\\
                \rR_{a'b'c'd}=&0\\
                \rR_{a'b'c'd'}=&R^{\esD'}_{a'b'c'd'}-2L_{a'b'}{}^{e}C_{ec'd'}
       }
}
From the final expressions of the Vidal curvature components the Schouten two-form and the Ricci tensor can be derived. For the Schouten two-form it is
easily seen that it ends up as a total exterior derivative of the two mean curvatures by looking at the trace of the curvature two-form in the
Cartan formalism, namely,
\eq{
	V:=R_{\bl{c}}{}^{\bl{c}}=\td\Tilde{\Tilde{\o}}_{\bl{c}}{}^{\bl{c}}=\td\k+\td\k'.
}
Now of course $\Tilde{\Tilde{\o}}$ is only a local object, and therefor it is not sure that $V$ can be written as an exact form globally - this leaves us with the following
proposition:
\prop{vidal schouten}{Let $\rdel$ be the Vidal connection then the Schouten two-form, $\rV$, of the Vidal connection can locally be written
	\eqal{1}{
		V=d\k_I
	}
	where $\k_I=\k+\k'$. From the integrability condition $dV=0$ it is obvious that
	\eqnono{
		V\in H^2(\ts),
	}
	where $H^2(\ts)$ denotes the second cohomology group of the manifold $\ts$.
	In component form the Schouten two-form looks like
	\eqal{1}{
		\rV_{ab}=&2(\rdel_{[a}\k')_{b]}-2L_{ab}{}^{c'}\k_{c'},\\
		\rV_{a'b}=&(\rdel_{a'}\k')_b-(\rdel_{b}\k)_{a'},\\
		\rV_{a'b'}=&2(\rdel_{[a'}\k)_{b']}-2L_{a'b'}{}^{c}\k'_{c}.\\
	}
}
Now finally the Ricci tensor and the curvature scalar of the Vidal connection can easily be derived.
\prop{vidal ricci}{Let $\rdel$ be the Vidal connection then the Ricci tensor reads in component form
	\eqal{1}{
		\rR_{ab}=&R^{\esD}_{ab}-2L_{ac}{}^{e'}C_{e'b}{}^c\\
		\rR_{a'b}=&-(\rdel_b\k)_{a'}+4L_{bc}{}^{e'}L'_{a'e'}{}^c\\
		\rR_{ab'}=&-(\rdel_{b'}\k')_{a}+4L_{ac}{}^{e'}L'_{b'e'}{}^c\\
		\rR_{a'b'}=&R^{\esD'}_{a'b'}-2L'_{a'c'}{}^{e}C_{eb'}{}^{c'}\\
	}
	and the Riemann curvature scalar is given by
	\eqnono{
		\rR=\Tilde{\Tilde{R}}+\Tilde{\Tilde{R}}'=R^{\esD}+R^{\esD'}-2L_{ab}{}^{c'}C_{c'}{}^{ab}-2L'_{a'b'}{}^{c}C_{c}{}^{a'b'}
	}
}
%
%
\subsection{The adapted connection}

In direct analogy to the previous section, several curvature- and torsion relations are derived
with respect to the adapted connection. In contrast to the Vidal-connection the adapted one 
is metric. In this case the Nijenhuis-tensor is related to the torsion in the way showed by the
next proposition. The metricity of the connection 
has its price though, as is seen below the torsion tensor is more complicated in this case. Some 
generalized Bianchi identities for the twisting tensor $L$ is also yielded in this case. All tensors, 
except the totally semi-basic ones, derived from the Riemann tensor, are expressed in terms 
of the irreducible parts of the deformation tensor.
\prop{ator}{Let the triplet $(\ts,\tg,I)$ define an riemannian almost product structure, let $N_I$ denote the Nijenhuis tensor of $I$ and 
	$\adel$ the adapted connection defined in \ref{adap}, then we have the following relation,
	\eqnono{
		{1\fr 2}N_I(X,Y)=\aT(X,Y)+\aT(IX,IY)
	}
}
\prop{adapted torsion}{Let $\aT$ be the torsion tensor of the adapted connection, $\adel$, then in component form it reads,
	\eqal{1}{
		\aT_{ab}{}^{c}=&0\\
		\aT_{ab}{}^{c'}=&-2L_{ab}{}^{c'}\\
		\aT_{a'b}{}^{c}=&-H_b{}^c{}_{a'}\\
		\aT_{ab'}{}^{c'}=&-H'_{b'}{}^{c'}{}_a\\
		\aT_{a'b'}{}^{c}=&-2L'_{a'b'}{}^{c}\\
		\aT_{a'b'}{}^{c'}=&0.
	}
	and
	\eqal{1}{
		\aT_a=&-\k'_a\\
		\aT_{a'}=&-\k_{a'}
	}
}
The curvature components can further be simplified as shown in the next proposition. Notice that they are expressed in terms of the Vidal connection.
\prop{adaptedcurvatures}{Let $\aR$ be the Riemann-tensor with respect to the adapted connection, then $\aR$ has the following components,
	\eqal{1}{
                \aR_{abc}{}^d=&\rR_{abc}{}^d+2L_{ab}{}^{e'}H_{c}{}^d{}_{e'}\\
                \aR_{abc}{}^{d'}=&0\\
                \aR_{abc'}{}^d=&0\\
                \aR_{abc'}{}^{d'}=&\rR_{abc'}{}^{d'}-2(\rdel_{[a}H')_{c'}{}^{d'}{}_{|b]}-2H'_{c'}{}^{e'}{}_{[a}H'_{e'}{}^{d'}{}_{|b]}\\
                \aR_{a'bc}{}^d=&\rR_{a'bc}{}^d+(\rdel_bH)_c{}^d{}_{a'}\\
                \aR_{a'bc}{}^{d'}=&0\\
                \aR_{a'bc'}{}^d=&0\\
                \aR_{a'bc'}{}^{d'}=&\rR_{a'bc'}{}^{d'}-(\rdel_{a'}H)_{c'}{}^{d'}{}_{b}\\
                \aR_{a'b'c}{}^d=&\rR_{a'b'c}{}^{d}-2(\rdel_{[a'}H)_{c}{}^{d}{}_{|b']}-2H_{c}{}^{e}{}_{[a'}H'_{e}{}^{d}{}_{|b']}\\
                \aR_{a'b'c}{}^{d'}=&0\\
                \aR_{a'b'c'}{}^d=&0\\
                \aR_{a'b'c'}{}^{d'}=&\rR_{a'b'c'}{}^{d'}+2L_{a'b'}{}^{e}H_{c'}{}^{d'}{}_e
        }
where,
	\eqal{1}{
		\aR_{abc'}{}^{d'}&:=E_a[\O_{bc'}{}^{d'}]-E_b[\O_{ac'}{}^{d'}]-C_{ab}{}^e\O_{ec'}{}^{d'}-2\O_{[a|c'}{}^{e'}\O_{b]e'}{}^{d'}
			-2L_{ab}{}^{e'}\o_{e'c'}{}^{d'}\\
		\aR_{a'b'c}{}^d&:=E_{a'}[\O_{b'c}{}^{d}]+E_{b'}[\O_{a'c}{}^{d}]-C_{a'b'}{}^{e'}\O_{e'c}{}^d-2\O_{[a'|c}{}^e\O_{b']e}{}^d
			-2L'_{a'b'}{}^e\o_{ec}{}^d
	}
}
Using proposition \ref{vidal curvature} the curvature tensor of the adapted connection can be written entirely in terms
of the semi-basic parts of the Vidal curvature, parts of the deformation tensors and the Vidal covariant derivative thereof.
\prop{adapted curvature vidal}{Let $\aR$ be the curvature tensor of the adapted connection then its components can be written
	\eqal{1}{
		\aR_{abcd}=&\rR_{abcd}+2L_{ab}{}^{e'} H_{cde'}\\
                \aR_{abcd'}=&0\\
                \aR_{abc'd}=&0\\
                \aR_{abc'd'}=&-2[(\rdel_{[a}L')_{c'd'|b]}+(\rdel_{[c'}L)_{ab|d']}-H'_{c'}{}^{e'}{}_{[a}H'_{d'e'|b]}]\\
                \aR_{a'bcd}=&(\rdel_bL)_{cda'}-2(\rdel_{[c}K)_{d]ba'}+2L_{cd}{}^{e'}L'_{a'e'b}-4L_{b[c}{}^{e'}L'_{a'e'|d]}\\
                \aR_{a'bcd'}=&0\\
                \aR_{a'bc'd}=&0\\
                \aR_{a'bc'd'}=&-(\rdel_{a'}L')_{c'd'b}+2(\rdel_{[c'}K')_{d']a'b}-2L'_{c'd'}{}^{e}L_{bea'}+4L'_{a'[c'}{}^{e}L_{be|d']}\\
                \aR_{a'b'cd}=&-2[(\rdel_{[a'}L)_{cd|b']}+(\rdel_{[c}L')_{a'b'|d]}-H_{c}{}^{e}{}_{[a'}H_{de|b']}]\\
                \aR_{a'b'cd'}=&0\\
                \aR_{a'b'c'd}=&0\\
                \aR_{a'b'c'd'}=&\rR_{a'b'c'd'}+2L_{a'b'}{}^{e}H'_{c'd'e}
       }
}
These relations can be expressed purely in terms of the adapted connection instead of the Vidal connection.
\prop{adapted curvature}{Let $\aR$ be the curvature tensor of the adapted connection then its components can be written
	\eqal{1}{
		\aR_{abcd}=&\rR_{abcd}+2L_{ab}{}^{e'} H_{cde'}\\
                \aR_{abcd'}=&0\\
                \aR_{abc'd}=&0\\
                \aR_{abc'd'}=&-2[(\adel_{[a}L')_{c'd'|b]}+(\adel_{[c'}L)_{ab|d']}-W'_{c'}{}^{e'}{}_{[a}W'_{d'e'|b]}+L'_{c'}{}^{e'}{}_{[a}L'_{d'e'|b]}+\\
			&+2L_{[a|}{}^{e}{}_{c'}L_{b]ed'}+2W_{[a}{}^{e}{}_{[c'}L_{b]e|d']}+{2\fr k}L_{ab[c'}\k_{d']}]\\
                \aR_{a'bcd}=&(\adel_bL)_{cda'}-2(\adel_{[c}K)_{d]ba'}-L_{cd}{}^{e'}H'_{e'a'b}-2K_{b[c|}{}^{e'}H'_{e'a'|d]}\\
                \aR_{a'bcd'}=&0\\
                \aR_{a'bc'd}=&0\\
                \aR_{a'bc'd'}=&-(\adel_{a'}L')_{c'd'b}+2(\adel_{[c'}K')_{d']a'b}+L_{c'd'}{}^{e}H'_{eba'}+2K_{a'[c'|}{}^{e}H'_{eb|d']}\\
                \aR_{a'b'cd}=&-2[(\adel_{[a'}L)_{cd|b']}+(\adel_{[c}L')_{a'b'|d]}-W_{c}{}^{e}{}_{[a'}W_{de|b']}+L_{c}{}^{e}{}_{[a'}L_{de|b']}+\\
			&+2L'_{[a'|}{}^{e'}{}_{c}L'_{b']e'd}+2W'_{[a'}{}^{e'}{}_{[c}L'_{b']e'|d]}+{2\fr k'}L'_{a'b'[c}\k'_{d]}]\\
                \aR_{a'b'cd'}=&0\\
                \aR_{a'b'c'd}=&0\\
                \aR_{a'b'c'd'}=&\rR_{a'b'c'd'}+2L_{a'b'}{}^{e}H'_{c'd'e}
       }
}
From previous proposition the Ricci tensor of the adpted connection can simply be deduced by contraction as the adapted 
connection is metric. It should be stressed that this Ricci tensor is in general not symmetric.
\prop{adapted ricci}{Let $\adel$ be the adapted connection then the Ricci tensor reads in component form
	\eqal{1}{
		\aR_{ab}=&\rR_{ab}+2L_{ac}{}^{e'}H_{b}{}^c{}_{e'}\\
		\aR_{a'b}=&(\adel_{c}H)_{b}{}^{c}{}_{a'}-H_{bc}{}^{e'}H'_{e'a'}{}^{c}-(\adel_b\k)_{a'}+\k^{e'}H'_{e'a'b}\\
		\aR_{ab'}=&(\adel_{c'}H')_{b'}{}^{c'}{}_{a}-H'_{b'c'}{}^{e}H_{ea}{}^{c'}-(\adel_{b'}\k)_{a}+\k'{}^{e}H_{eab'}\\
		\aR_{a'b'}=&\rR_{a'b'}+2L'_{a'c'}{}^{e}H'_{b'}{}^{c'}{}_e\\
	}
	and the Riemann curvature scalar is given by
	\eqnono{
		\aR=\Tilde{R}+\Tilde{R}'=\Tilde{\Tilde{R}}+\Tilde{\Tilde{R}}'+2L_{ab}{}^{c'}L^{ab}{}_{c'}+2L'_{a'b'}{}^{c}{L'}^{a'b'}{}_{c}
	}
}
The generalized Biachi-identities for the twisting tensor $L$ are given in next proposition. 
They were derived by using the antisymmetry ((i) of proposition \ref{curvatureidentities}) and the components
expressed in terms of the adapted connection, see proposition \ref{adapted curvature}.
\prop{twistbianchi2}{Let $\adel$ be the adapted connection associated with an almost product structure and $L,L',K,K'$ be the respective twisting and
       extrinsic curvature tensors of the
       associated distributions then the following identities hold
\eqal{1}{
                \adel_{[a}L_{bc]}{}^{d'}+L_{[ab|}{}^{e'}H_{e'}{}^{d'}{}_{|c]}=&0\\
                \adel_{[a'}L_{b'c']}{}^{d}+L_{[a'b'|}{}^{e}H_{e}{}^{d}{}_{|c']}=&0
                }
}
As was seen in previous subsection, where the Vidal connection was studied, new identities between parts of the deformation tensors
arose as integrability conditions on these while imposing the identities of the curvature tensor. 
These were derived in proposition \ref{adapted curvature} above and in the same fashion the corresponding identities for
the adapted connection arise.
\prop{adepted-identity}{Let $\adel$ be the adapted connection, $K,L,K',L'$ be the second fundamental tensors with respect to a almost product
	structure, $I$, then the following identities hold
	\eqal{1}{
		(\adel_{[a}K')_{c'd'|b]}+(\adel_{(c'}L)_{ab|d')}-2L'_{(c'|}{}^{e'}{}_{[a}K'_{e'|d')|b]}-2K_{[a|}{}^{e}{}_{(c'}L_{e|b]d')}=&0,\\
		(\adel_{[a'}K)_{cd|b']}+(\adel_{(c}L')_{a'b'|d)}-2L_{(c|}{}^{e}{}_{[a'}K_{e|d)|b']}-2K'_{[a'|}{}^{e'}{}_{(c}L'_{e'|b']d)}=&0.
	}
	These identities look in the contracted case like
	\eqal{1}{
		(\adel_{[a}\k')_{b]}+(\adel_{c'}L)_{ab}{}^{c'}-2W_{[a|}{}^{e}{}_{c'}L_{e|b]}{}^{c'}-{2\fr k}L_{ab}{}^{c'}\k_{c'}=&0,\\
		(\adel_{[a'}\k)_{b']}+(\adel_{c}L)_{a'b'}{}^{c}-2W_{[a'|}{}^{e'}{}_{c}L_{e'|b']}{}^{c}-{2\fr k'}L_{a'b'}{}^{c}\k'_{c}=&0.
	}
}
%
%
\subsection{The Levi-Civita connection}

In this section all curvature relations for the Levi-Civita connection is given. The curvature, Ricci
and the curvature scalar are  expressed in terms of the irreducible components of the deformation-tensor. 
Starting from Cartan's structure equations and writing the curvature two-form as,
\eqal{1}{
	\tR_{\bc}{}^{\bd}:=&\td \ul{\o}_{\bc}{}^{\bd}-\ul{\o}_{\bc}{}^{\bl{e}}\wedge \ul{\o}_{\bl{e}}{}^{\bd}=\\
			=&\mx{cc}{\td\ul{\o}_{c}{}^d-\ul{\o}_{c}{}^{e}\wedge \ul{\o}_{e}{}^{d}-\ul{H}_{c}{}^{e'}\wedge \ul{H}_{e'}{}^{d},&
					\td \ul{H}_{c}{}^{d'}-\ul{\o}_{c}{}^{e}\wedge \ul{H}_{e}{}^{d'}-H_{c}{}^{e'}\wedge \ul{\o}_{e'}{}^{d'}\\
			\td \ul{H}_{c'}{}^{d}-\ul{\o}_{c'}{}^{e'}\wedge \ul{H}_{e'}{}^{d}-H_{c'}{}^{e}\wedge \ul{\o}_{e}{}^{d},&
				\td\ul{\o}_{c'}{}^{d'}-\ul{\o}_{c'}{}^{e'}\wedge \ul{\o}_{e'}{}^{d'}-\ul{H}_{c'}{}^{e}\wedge \ul{H}_{e}{}^{d'}}
}
the components can be given in terms of the adapted connection.
\prop{levcivcurvatures}{Let $\tR$ be the Riemann-tensor with respect to the Levi-Civita connection, then $\tR$ has the following components,
\eqal{1}{
        \tR_{abc}{}^d=&\aR_{abc}{}^d+2H_{[a|c}{}^{e'}H_{b]}{}^{d}{}_{e'}\\
        \tR_{abc}{}^{d'}=&2(\adel_{[a}H)_{b]c}{}^{d'}+2L_{ab}{}^{e'}H'_{e'}{}^{d'}{}_c\\
        \tR_{abc'}{}^d=&-2(\adel_{[a}H)_{b]}{}^{d}{}_{c'}-2L_{ab}{}^{e'}H'_{e'c'}{}^{d}\\
        \tR_{abc'}{}^{d'}=&\aR_{abc'}{}^{d'}+2H_{[a|}{}^{e}{}_{c'}H_{b]e}{}^{d'}\\
        \tR_{a'bc}{}^d=&\aR_{a'bc}{}^{d}-H'_{a'}{}^{e'}{}_cH_{b}{}^d{}_{e'}+H_{bc}{}^{e'}H'_{a'e'}{}^d\\
        \tR_{a'bc}{}^{d'}=&(\adel_{a'}H)_{bc}{}^{d'}+(\adel_{b}H')_{a'}{}^{d'}{}_c-H_b{}^e{}_{a'}H_{ec}{}^{d'}-H'_{a'}{}^{e'}{}_bH'_{e'}{}^{d'}{}_c\\
        \tR_{a'bc'}{}^d=&-(\adel_{a'}H)_{b}{}^{d}{}_{c'}-(\adel_{b}H')_{a'c'}{}^{d}+H_{b}{}^{e}{}_{a'}H_{e}{}^{d}{}_{c'}+H'_{a'}{}^{e'}{}_{b}H'_{e'c'}{}^{d}\\
	\tR_{a'bc'}{}^{d'}=&\aR_{a'bc'}{}^{d'}-H'_{a'c'}{}^{e}H_{be}{}^{d'}+H_{b}{}^{e}{}_{c'}H'_{a'}{}^{d'}{}_e\\
        \tR_{a'b'c}{}^d=&\aR_{a'b'c}{}^{d}+2H'_{[a'|}{}^{e'}{}_cH'_{b']e'}{}^{d}\\
        \tR_{a'b'c}{}^{d'}=&-2(\adel_{[a'}H')_{b']}{}^{d'}{}_c-2L'_{a'b'}{}^{e}H_{ec}{}^{d'}\\
        \tR_{a'b'c'}{}^d=&2(\adel_{[a'}H')_{b']c'}{}^{d}+2L'_{a'b'}{}^{e}H_{e}{}^{d}{}_{c'}\\
        \tR_{a'b'c'}{}^{d'}=&\aR_{a'b'c'}{}^{d'}+2H'_{[a'|c'}{}^eH'_{b']}{}^{d'}{}_e\\
      }
}
These are also known as the Gauss-Codazzi relations. In the case of the Levi--Civita connection though, it is clear that its curvature
tensor possesses the box symmetry, i.e. $$\yng(2,2)$$
From the Tic--Tac--Toe notation it follows that the box symmetry reduces to six irreducible parts under a rigging. From the previous
analysis of the Vidal and the adapted curvatures these can again be written entirely in terms of the semi-basic components and parts of
the deformation tensors.
\prop{leviadapt}{Let $\tR$ be the Riemann tensor with respect to the Levi--Civita connection then its components can be written in terms of just
	the deformation tensor and adapted covariant derivatives thereof plus the complete longitudal and normal parts of the adapted curvature,
	\eqal{1}{
		\tR_{abcd}=&\aR_{abcd}+2H_{[a|c}{}^{e'}H_{b]de'}\\
		\tR_{abcd'}=&2(\adel_{[a}H)_{b]cd'}+2L_{ab}{}^{e'}H'_{e'd'c}\\
		\tR_{abc'd'}=&-2[(\adel_{[c'}L)_{ab|d']}+(\adel_{[a}L')_{c'd'|b]}+L_{[a|}{}^e{}_{c'}L_{b]ed'}+L'_{[c'|}{}^{e'}{}_{a}L'_{d']e'b}-\\
			&-W_{[a|}{}^e{}_{c'}W_{b]ed'}-W'_{[c'|}{}^{e'}{}_{a}W'_{d']e'b}]\\
	\tR_{a'bc'd}=&\frac{1}{2}\tR_{a'c'bd}-(\adel_{(a'|}K)_{bd|c')}-(\adel_{(b}K')_{a'c'|d)}+K_{(b|}{}^e{}_{a'}K_{d)ec'}-L_{(b|}{}^e{}_{a'}L_{d)ec'}+\\
                             &+K'_{(a'|}{}^{e'}{}_{b}K'_{c)e'd}-L'_{(a'|}{}^{e'}{}_{b}L'_{c)e'd}\\
		\tR_{a'b'c'd}=&2(\adel_{[a'}H')_{b']c'd}+2L'_{a'b'}{}^{e}H_{edc'}\\
		\tR_{a'b'c'd'}=&\aR_{a'b'c'd'}+2H'_{[a'|c'}{}^{e}H'_{b']d'e}\\
	}
}	
In ref. \cite{Gr67} these were deduced in terms of the Levi--Civita connection but as it does not preserve the rigging these are
better expressed in terms of the adapted or the Vidal connection. Note that the expressions are only decomposed in terms of the irreducible
parts of the deformation tensors, 
where it is necessary in order to make manifest the symmetries. In all other cases it is a straight forward process to do just by insertion. 
\prop{levividal}{Let $\tR$ be the Riemann tensor with respect to the Levi--Civita connection then its components can be written in terms of just
	the deformation tensor and Vidal covariant derivatives thereof plus the complete longitudal and normal parts of the Vidal curvature as
	\eqal{1}{
		\tR_{abcd}=&\rR_{abcd}+2L_{ab}{}^{e'}H_{cde'}+2H_{[a|c}{}^{e'}H_{b]de'}\\
		\tR_{abcd'}=&2(\rdel_{[a}H)_{b]cd'}+2H_{[b|c}{}^{e'}H'_{d'e'|a]}+2L_{ab}{}^{e'}H'_{e'd'c}\\
		\tR_{abc'd'}=&-2[(\rdel_{[c'}L)_{ab|d']}+(\rdel_{[a}L')_{c'd'|b]}+{2\fr k'}\k'_{[a}L'_{c'd'|b]}+{2\fr k}\k_{[c'}L_{ab|d']}\\
		&-W_{[a|}{}^e{}_{c'}W_{b]ed'}-W'_{[c'|}{}^{e'}{}_{a}W'_{d']e'b}-2W_{[a|}{}^e{}_{[c'}L_{b]e|d']}-2W'_{[c'|}{}^{e'}{}_{[a}L'_{d']e'|b]}-\\
			&-L_{[a|}{}^e{}_{c'}L_{b]ed'}-L'_{[c'|}{}^{e'}{}_{a}L'_{d']e'b}]\\
		\tR_{a'bc'd}=&\frac{1}{2}\tR_{bda'c'}-\rdel_{(a'|}K_{bd|c')}-\rdel_{(b|}
                	K'_{a'c'|d)}-W_{(d|}{}^{e}{}_{(a'}W_{b)ec')}-
                	L_{(d|}{}^{e}{}_{(a'}L_{b)ec')}-\\&-2W_{(d}{}^{e}{}_{(a'|}L_{b)e|c')}
			-\frac{1}{k^2}\k_{a'}\k_{c'}\eta_{bd}-\frac{1}{k}\k_{(a'|}W_{bd|c')}-\\
			&-W'_{(c'|}{}^{e'}{}_{(b|}W'_{a')e'd)}-
                	L'_{(c'|}{}^{e'}{}_{(b}L'_{d)e'a')}-\\&-2W'_{(c'}{}^{e'}{}_{(b|}L'
                       _{d)e'|a')}-\frac{1}{{k'}^2}\k'_{b}\k'_{d}\eta_{a'c'}
                       -\frac{1}{k'}\k'_{(b|}W_{a'c'|d)}\\
		\tR_{a'b'c'd}=&2(\rdel_{[a'}H')_{b']c'd}+2H'_{[b'|c'}{}^{e}H_{de|a']}+2L'_{a'b'}{}^{e}H_{edc'}\\
		\tR_{a'b'c'd'}=&\rR_{a'b'c'd'}+2L'_{a'b'}{}^{e}H'_{c'd'e}+2H'_{[a'|c'}{}^{e}H'_{b']d'e}\\
	}
}
From propositions \ref{leviadapt} and \ref{levividal} the Ricci tensor and the curvature scalar is most easily deduced.
\prop{Ricciadapted}{Let $\tdel$ be the Levi-Civita connection then the Ricci tensor reads in terms of the adapted connection,
	\eqal{1}{
		\tR_{ab}=&R_{ab}+R''_{ab}=\aR_{(ab)}-\frac{1}{k}(\adel_{c'}{\k})^{c'}{\eta}_{ab}-(\adel_{(a}{\k}')_{b)}-(\adel_{c'}W)_{ab}{}^{c'}
			+\frac{1}{k}{\k}^2{\eta}_{ab}+\\
			&+W_{ab}{}^{c'}\k_{c'}+\frac{1}{k'}{\k'}_{a}{\k'}_b+W'_{c'e'a}{W'}^{c'e'}{}_b-L'_{c'e'a}{L'}^{c'e'}{}_b\\
		\tR_{ab'}=&(\frac{1-k}{k})({\adel}_{a}\k)_{b'}+(\frac{1-k'}{k'})({\adel}_{b'}\k')_a
			+(\adel_cW)_a{}^c{}_{b'}+\\
			&+(\adel_{c'}W')_{b'}{}^{c'}{}_a+(\adel_cL)_a{}^c{}_{b'}+(\adel_{c'}L')_{b'}{}^{c'}{}_a+\\
			&+4L_{a}{}^{ec'}L'_{b'c'e}-2L_{a}{}^{ce'}{W'}_{b'e'c}-2L'_{b'}{}^{c'e}{W}_{aec'}-\frac{2}{k'}L_{acb'}{\k'}^{c}
				-\frac{2}{k}L_{b'c'a}{\k}^{c'},\\
		\tR_{a'b'}=&R'_{a'b'}+R''_{a'b'}=\aR_{(a'b')}-\frac{1}{k'}(\adel_{c}{\k'})^{c}{\eta'}_{a'b'}-(\adel_{(a'}{\k})_{b')}
			-(\adel_{c}W')_{a'b'}{}^{c}+\frac{1}{k'}{\k'}^2{\eta'}_{a'b'}+\\
			&+W'_{a'b'}{}^{c}\k'_{c}+\frac{1}{k}{\k}_{a'}{\k}_{b'}+W_{cea'}{W}^{ce}{}_{b'}-L_{cea'}{L}^{ce}{}_{b'}
	}
	where the following definitions are used
	\eqal{1}{
		R_{ab}:=&\tR_{acb}{}^c=\aR_{(ab)}-W_{a}{}^{ec'}W_{bec'}+(\frac{k-2}{k})\k^{c'}W_{abc'}+(\frac{k-1}{k^2})
			\eta_{ab}{\k}^2+L_{a}{}^{ec'}L_{bec'}\\
		R''_{ab}:=&\tR_{ac'b}{}^{c'}=-(\adel_{(a}\k')_{b)}-(\adel_{c'}W)_{ab}{}^{c'}
			-\frac{1}{k}(\adel_{c'}\k)^{c'}\eta_{ab}+{W'}^{c'e'}{}_a{W'}_{c'e'b}+
			\frac{1}{k'}\k'_a\k'_b-\\&-{L'}^{c'e'}{}_aL'_{c'}{}_{e'b}
                         +W_{a}{}^{ec'}W_{bec'}+\frac{2}{k}\k^{c'}W_{abc'}+\frac{1}{k^2}
			\eta_{ab}{\k}^2-L_{a}{}^{ec'}L_{bec'}
        }
	and similarly for the $R'_{a'b'}$ component.
	The Riemann curvature scalar is given by
	\eqal{1}{
		\tR=&R+2R''+R'=\Tilde{R}+\Tilde{R}'+\frac{1-k}{k}\k^2+\frac{1-k'}{k'}{\k'}^2-2\tdel\cdot\k_I+
                    W^2+{W'}^2-L^2-{L'}^2
	}
	where the following definitions are used
	\eqal{1}{
		R:=&\tR_{ab}{}^{ab}=\Tilde{R}+{\k}^2-{K}^2+{L}^2\\
                R'':=&\tR_{ab'}{}^{ab'}=-\adel_{a'}\k^{a'}-\adel_{a}{\k'}^a+W^2-L^2+\frac{1}{k}{\k}^2+{W'}^2-{L'}^2+\frac{1}{k'}{\k'}^2\\
                R':=&\tR_{a'b'}{}^{a'b'}=\Tilde{R}'+{\k'}^2-{K'}^2+{L'}^2\\
	}
}
Part of which is found in \cite{Roc84}. In the expressions on the curvature scalar above, we
have used the following relation
\eqs{
	\tdel\cdot\k_I=(\adel_{a'}\k)^{a'}+(\adel_{a}\k')^a-{\k}^2-{\k'}^2\\
}
and the notation $L^2:=L_{abc'}L^{abc'}$ etc. In terms of the Vidal connection the Ricci tensor and the curvature scalar is given in next proposition.
\prop{Riccividal}{Let $\tdel$ be the Levi-Civita connection then the Ricci tensor reads, in component form, in terms of the Vidal connection
	\eqal{1}{
		\tR_{ab}=&\aR_{(ab)}-\frac{1}{k}(\rdel_{c'}{\k})^{c'}{\eta}_{ab}
              		-(\rdel_{(a}{\k}')_{b)}-(\rdel_{c'}W)_{ab}{}^{c'}-2W_a{}^{ec'}W_{ebc'}+\\
			&+(\frac{k-2}{k})W_{ab}{}^{c'}\k_{c'}
			+\frac{1}{k'}{\k'}_{a}{\k'}_b+\frac{1}{k}{\k}^2{\eta}_{ab}+W'_{c'e'a}{W'}^{c'e'}{}_b-\\
				&-L'_{c'e'a}{L'}^{c'e'}{}_b+2L_{a}{}^{ec'}L_{bec'}\\
		\tR_{ab'}=&(\frac{1-k}{k})({\rdel}_{a}\k)_{b'}+(\frac{1-k'}{k'})({\rdel}_{b'}\k')_a
			+(\rdel_cW)_a{}^c{}_{b'}+\\
			&+(\rdel_{c'}W')_{b'}{}^{c'}{}_a+(\rdel_cL)_a{}^c{}_{b'}+(\rdel_{c'}L')_{b'}{}^{c'}{}_a+\\
			&+6L_{a}{}^{ec'}L'_{b'c'e}+2W_{a}{}^{ec'}W'_{b'c'e}-(\frac{k+k'-2}{kk'})\k'_a\k_{b'}-\\
				&-L_{acb'}{\k'}^{c}
				-L_{b'c'a}{\k}^{c'}-(\frac{k'-2}{k'})W_{acb'}{\k'}^{c}
				-(\frac{k-2}{k})W_{b'c'a}{\k}^{c'}\\
		\tR_{a'b'}=&\aR_{(a'b')}-\frac{1}{k'}(\rdel_{c}{\k'})^{c}{\eta'}_{a'b'}-(\rdel_{(a'}{\k})_{b')}-(\rdel_{c}W')_{a'b'}{}^{c}
			2W'_{a'}{}^{e'c}W'_{e'b'c}-\\&+(\frac{k'-2}{k'})W'_{a'b'}{}^{c}\k'_{c}
			+\frac{1}{k}{\k}_{a'}{\k}_{b'}+\frac{1}{k'}{\k'}^2{\eta'}_{a'b'}+W_{cea'}{W}^{ce}{}_{b'}-\\
			&-L_{cea'}{L}^{ce}{}_{b'}+2L'_{a'}{}^{e'c}L'_{e'b'c}
	}
	where the following definitions were used
	\eqal{1}{
		R_{ab}:=&\tR_{acb}{}^c=\rR_{(ab)}-W_{a}{}^{ec'}W_{bec'}+(\frac{k-2}{k})\k^{c'}W_{abc'}+(\frac{k-1}{k^2})
			\eta_{ab}{\k}^2\\
			&+3L_{a}{}^{ec'}L_{bec'}+2L_{(a|}{}^{ce'}W_{b)ce'}\\
		R''_{ab}:=&\tR_{ac'b}{}^{c'}=-(\rdel_{(a}\k')_{b)}-(\rdel_{c'}W)_{ab}{}^{c'}
			-\frac{1}{k}(\rdel_{c'}\k)^{c'}\eta_{ab}+{W'}^{c'e'}{}_a{W'}_{c'e'b}+
			\frac{1}{k'}\k'_a\k'_b-\\&-{L'}^{c'e'}{}_aL'_{c'}{}_{e'b}
                         -W_{a}{}^{ec'}W_{bec'}+\frac{1}{k^2}
			\eta_{ab}{\k}^2-L_{a}{}^{ec'}L_{bec'}-2L_{(a}{}^{ec'}W_{e|b)c'}
	}
	again similarly for the $R'_{a'b'}$ component.
	The Riemann curvature scalar is given by
	\eqal{1}{
		\tR=&R+2R''+R'=\Tilde{\Tilde{R}}+\Tilde{\Tilde{R'}}+\frac{1-k}{k}\k^2+\frac{1-k'}{k'}{\k'}^2-2\tdel\cdot\k_I+W^2+{W'}^2+L^2+{L'}^2
	}
	where the following definitions were used
	\eqal{1}{
		R=&\Tilde{\Tilde{R}}+\k^2-K^2+3L^2\\
                R''=&-\rdel_{a'}\k^{a'}-\rdel_{a}{\k'}^a+W^2-L^2+\frac{1}{k}{\k}^2+{W'}^2-{L'}^2+\frac{1}{k'}{\k'}^2\\
                R'=&\Tilde{\Tilde{R'}}+{\k'}^2-{K'}^2+3{L'}^2
        }        
}
%
%
\subsection{Their conformal properties}

In an earlier treatment it was found that conformal transformations does affect a connection including torsion and non-metricity in a
non-trivial fashion. Here will be given a complete analysis of the induced transformations of the Vidal, adapted and Levi--Civita connections.

\defn{Conform trans of con}{Let the triplet $(\ts,\tg,I)$ denote an almost product manifold, let $\tdel,\adel,\rdel$ denote the Levi--Civita,
	adapted and Vidal connection respectively then define the associated {\bf conformal tensors} denoted $\tsC,\aC,\rC$ with characteristics
	\eqal{1}{
		\tsC,\aC,\rC:&\qquad \L^1\times\L^1\longmapsto \L^1
	}
	by
	\eqal{1}{
		\tsC(X,Y):=&\ct \tdel_XY-\tdel_XY\\
		\aC(X,Y):=&\ct \adel_XY-\adel_XY\\
		\rC(X,Y):=&\ct \rdel_XY-\rdel_XY\\
	}
	where $X,Y\in L^1$ are vectorfields on $\ts$.
}
The conformal tensor, corresponding to the Vidal and the adapted connection, can most easily be expressed in terms of the conformal tensor of the
Levi--Civita connection.
\prop{conf trans}{Let $\tsC,\aC,\rC$ be the conformal tensors defined in \ref{Conform trans of con} then following relations hold
	\eqal{1}{
		\tsC(X,Y)&=X[\phi]Y+Y[\phi]X-\tg(X,Y)\ri\td\phi\\
		\aC(X,Y)&=\ico \tsC(X,\ico Y)+\nco \tsC(X,\nco Y)\\
		\rC(X,Y)&=\ico \tsC(\ico X,\ico Y)+\nco \tsC(\nco X,\nco Y)\\
	}
	where $X,Y\in \L^1$ are vectorfields on $\ts$.
}
\proof{To be added.}
The difference between these conformal tensors manifests itself in a clearer way by studying the expressions in component form. The conformal
tensor of the Levi--Civita connection is read of from the above proposition and is noticeably symmetric. 
\prop{levi conf}{Let $\tsC$ be the conformal tensor of the Levi--Civita connection then in component form it reads,
	\eqal{1}{
		\tsC_{ab}{}^{c}=&2\d_{(a}^cE_{b)}[\phi]-\eta_{ab}\eta^{cd}E_d[\phi]\\
		\tsC_{ab}{}^{c'}=&-\eta_{ab}\eta^{c'd'}E_{d'}[\phi]\\
		\tsC_{a'b}{}^{c}=&E_{a'}[\phi]\d_{b}^c\\
		\tsC_{ab'}{}^{c'}=&E_{a}[\phi]\d_{b'}^{c'},\\
		\tsC_{a'b'}{}^{c}=&-\eta_{a'b'}\eta^{cd}E_{d}[\phi]\\
		\tsC_{a'b'}{}^{c'}=&2\d_{(a'}^{c'}E_{b')}[\phi]-\eta_{a'b'}\eta^{c'd'}E_{d'}[\phi]\\
	}
}
The conformal tensor of the adapted connection can now be derived from the above expressions. It should be stressed though that it is not symmetric.
\prop{adapt conf}{Let $\aC$ be the conformal tensor of the adapted connection then in component form it reads
	\eqal{1}{
		\aC_{ab}{}^{c}=&2\d_{(a}^cE_{b)}[\phi]-\eta_{ab}\eta^{cd}E_d[\phi]\\
		\aC_{ab}{}^{c'}=&0\\
		\aC_{ab'}{}^{c}=&0\\
		\aC_{a'b}{}^{c}=&E_{a'}[\phi]\d_{b}^c\\
		\aC_{ab'}{}^{c'}=&E_{a}[\phi]\d_{b'}^{c'}\\
		\aC_{a'b}{}^{c'}=&0\\
		\aC_{a'b'}{}^{c}=&0\\
		\aC_{a'b'}{}^{c'}=&2\d_{(a'}^{c'}E_{b')}[\phi]-\eta_{a'b'}\eta^{c'd'}E_{d'}[\phi]\\
	}
}
Finally in the case of the Vidal connection the conformal tensor takes a very simple form and will, like in the Levi--Civita case, be symmetric.
\prop{vidal conf}{Let $\rC$ be the conformal tensor of the Vidal connection then in component form it reads
	\eqal{1}{
		\rC_{ab}{}^{c}=&2\d_{(a}^cE_{b)}[\phi]-\eta_{ab}\eta^{cd}E_d[\phi]\\
		\rC_{ab}{}^{c'}=&0\\
		\rC_{ab'}{}^{c}=&0\\
		\rC_{a'b}{}^{c}=&0\\
		\rC_{ab'}{}^{c'}=&0\\
		\rC_{a'b}{}^{c'}=&0\\
		\rC_{a'b'}{}^{c}=&0\\
		\rC_{a'b'}{}^{c'}=&2\d_{(a'}^{c'}E_{b')}[\phi]-\eta_{a'b'}\eta^{c'd'}E_{d'}[\phi]\\
	}
}
Well known is the fact that when decomposing the Riemann curvature tensor of the Levi--Civita connection into its irreducible parts with
respect to its traces, the appearing Weyl tensors measures whether the riemannian manifold is conformally flat or not. More specifically, the vanishing
of the Weyl tensor is the condition for local conformal flatness in the case when the dimension of the manifold exceeds three. The tracefree part of the Ricci tensor is defined by,
\eq{
        \hat{\tR}_{\ba\bb}:=\tR_{\ba\bb}-{1\fr m}\tR\eta_{\ba\bb}
}
Next the Ricci one-form and the tracefree Ricci one-form are defined.
\defn{riccioneform}{Let $\tR_{ab},\hat{\tR}_{ab}$ denote the Ricci tensor and its tracefree part respectively then define their one forms by
        \eqal{1}{
                 \tR^\ba:=&E^\bb\tR_\bb{}^\ba\\
                 \hat{\tR}^\ba:=&E^\bb\hat{\tR}_\bb{}^\ba=\tR^\ba-{1\fr m}\tR E^\ba\\
        }
}
Denoting the Rimann two form $\tR^{\tc\td}={1\fr 2}E^\ta\wedge E^\tb\tR_{\ta\tb}{}^{\tc\td}$ and the Weyl two form by 
$\tC^{\tc\td}={1\fr 2}E^\ta\wedge E^\tb\tC_{\ta\tb}{}^{\tc\td}$
the decomposition is most elegantly written
\eqnono{
	\arr{ccccccc}{
	\yng(2,2)		&=&	\widetilde{\yng(2,2)}		&\oplus&	\widetilde{\yng(2)}	&\oplus&	\bigodot\vspace{.3cm}\\
	{1\fr 12}m^2(m^2-1)	&=&	{1\fr 12}(m-3)m(m+1)(m+2)	&+&		{1\fr 2}(m-1)(m+2)	&+&		1\vspace{.3cm}\\
	\tR^{\bc\bd}		&=&	\tC^{\bc\bd}			&+&{2\fr m-2}\hat{\tR}^{[\bc}\wedge E^{\bd]}&+&	{1\fr {m(m-1)}}\tR E^\bc\wedge E^\bd
	}
}
After reinserting the expression for the tracefree part of the Ricci tensor and solving for the Weyl tensor the more familiar form is obtained
\eq{
        \tC_{\ba\bb}{}^{\bc\bd}=\tR_{\ba\bb}{}^{\bc\bd}-{4\fr m-2}\d_{[\ba}^{[\bc}\tR_{\bb]}^{\bd]}+{2\fr (m-1)(m-2)}\tR\d_{[\ba}^{[\bc}\d_{\bb]}^{\bd]}
}
From this equation the different components are read off and put in next proposition.
\prop{weyl}{Let $\tC^{\bc\bd}$ be the weyl tensor in an almost product manifold then its components look like
        \eqal{1}{
                 \tC_{ab}{}^{cd}=&\tR_{ab}{}^{cd}-{4\fr m-2}\d_{[a}^{[c}\tR_{b]}^{d]}+{2\fr (m-1)(m-2)}\tR\d_{[a}^{[c}\d_{b]}^{d]}\\
                 \tC_{ab}{}^{cd'}=&\tR_{ab}{}^{cd'}-{2\fr m-2}\d_{[a}^{c}\tR_{b]}^{d'}\\
                 \tC_{ab}{}^{c'd'}=&\tR_{ab}{}^{c'd'}\\
                 \tC_{a'b}{}^{c'd}=&\tR_{a'b}{}^{c'd}-{1\fr m-2}(\d_{a'}^{c'}\tR_{b}^{d}+\d_{b}^{d}\tR_{a'}^{c'})+{1\fr (m-1)(m-2)}\tR\d_{a'}^{c'}\d_{b}^{d}\\
                 \tC_{a'b'}{}^{c'd}=&\tR_{a'b'}{}^{c'd}-{2\fr m-2}\d_{[a'}^{c'}\tR_{b']}^{d}\\
                 \tC_{a'b'}{}^{c'd'}=&\tR_{a'b'}{}^{c'd'}-{4\fr m-2}\d_{[a'}^{[c'}\tR_{b']}^{d']}+{2\fr (m-1)(m-2)}\tR\d_{[a'}^{[c'}\d_{b']}^{d']}\\
        }
}
From above proposition it is for instance clear that the $\tR_{abc'}{}^{d'}$ component of the Riemann tensor must be conformally invariant. Taking a
look at its final expression in proposition \ref{leviadapt} the only non-manifest conformally invariant terms, are those involving derivatives. These can
be proved to be independently conformally invariant. 
\prop{derlconf}{Let $L,L'$ be the respective twisting tensors of the characteristic distributions defined by an almost product structure on $\ts$ then
        the following relations hold
        \eqal{1}{
                \ct\adel_{[Z'}L_{W']}(X,Y)=&e^{2\phi}\adel_{[Z'}L_{W']}(X,Y)\\
                \ct\adel_{[Z}L'_{W]}(X',Y')=&e^{2\phi}\adel_{[Z}L'_{W]}(X',Y')
        }
        or put in component form
        \eqal{1}{
                (\ct\adel_{[c'}L)_{ab|d']}=&e^{2\phi}(\adel_{[c'}L)_{ab|d']}\\
                (\ct\adel_{[c}L')_{a'b'|d]}=&e^{2\phi}(\adel_{[c}L')_{a'b'|d]}
        } 
}
Proceeding in the same fashion as in \cite{Carter},investigating how the curvature components of the adapted connection can be divided into irreducible parts - 
in the case of an almost product manifold instead as for just an embedding, some immediate differences is noticed. The generalisation of what Carter calls the
outer curvature is the $\aR_{ab}{}^{c'd'}$ component which will be seen not to be conformally invariant in the generic case but only if the unprimed
distribution is integrable. The same is of course true for the outer curvature of the complementary (primed) distribution.
From proposition \ref{adapted curvature} it is manifest that the non-invariant components are $${2\fr k}L_{ab[c'}\k_{d']}$$ and $${2\fr k'}L'_{a'b'[c}\k'_{d]}$$
respectively. For the internal curvature which in the language of almost product manifolds is the total semi-basic components of the adapted curvature,
one can follow the procedure of dividing the tensor components into its irreducible parts according to the above scheme. In the generic case
the semi-basic components does not have the box symmetry $$\yng(2,2)$$. Because of torsion though, it has the other symmetry parts. It is clear
that when the  distribution is integrable its internal curvature will indeed have the box symmetry. Defining the internal Weyl tensors
the generalization of \cite{Carter,Carter97} can be made.
\defn{intweyl}{Let $\aR$ be the curvature tensor of the adapted connection and $\aR_{ab}{}^{cd}$, $\aR_{a'b'}{}^{c'd'}$ the internal curvatures of
        the two complementary distributions associated with an almost product manifold then define their respective Weyl tensors by
        \eqal{1}{
                \tilde{C}_{ab}{}^{cd}=&\tilde{R}_{ab}{}^{cd}-{4\fr k-2}\d_{[a}^{[c}\tilde{R}_{b]}^{d]}+{2\fr (k-1)(k-2)}\tilde{R}\d_{[a}^{[c}\d_{b]}^{d]}\\
                \tilde{C}'_{a'b'}{}^{c'd'}=&{{\tilde{R}}'}_{a'b'}{}^{c'd'}-{4\fr k'-2}\d_{[a'}^{[c'}\tilde{R}'{}_{b']}^{{d'}]}
                       +{2\fr (k'-1)(k'-2)}{\tilde{R}'}\d_{[a'}^{[c'}\d_{b']}^{d']}\\
        }
        where
        \eqal{3}{
               \tilde{R}_{ab}{}^{cd}:=&\aR_{ab}{}^{cd},& \tilde{R}_{a}^{b}:=&\aR_{ac}{}^{bc},& \tilde{R}:=&\aR_{ab}{}^{ab},\\
        \tilde{R}'{}_{a'b'}{}^{c'd'}:=&\aR_{a'b'}{}^{c'd'},&\qquad \tilde{R}'{}_{a'}^{b'}:=&\aR_{a'c'}{}^{b'c'},&\qquad \tilde{R}':=&\aR_{a'b'}{}^{a'b'}.
       }
}
Following Carter's procedure it is now easy to generalize his relations to the case of an almost product manifold.
\prop{weyl-weyl}{Let $\tilde{C},\tilde{C}'$ be the Weyl tensors of the internal curvatures of the two complemenatry distributions associated with an
         almost product manifold. Then the following relations hold.
\eqal{1}{
	\tilde{C}_{ab}{}^{cd}=&C_{ab}{}^{cd}-\frac{4}{k-2}\d^{[c}_{[a}C^{d]}_{b]}+\frac{2}{(k-1)(k-2)}C-\\
                  &- 2(L_{[a|}{}^c{}_{e'}L_{b]}{}^{de'}+2L_{[a|}{}^{[c}{}_{e'}W_{b]}{}^{d]e'})-\\
                  & -\frac{4}{k-2}(\d_{[a|}^{[c}L_{b]}{}^e{}_{e'}L_e{}^{d]e'}+
                   \d_{[a|}^{[c}W_{b]}{}^e{}_{e'}W_e{}^{d]e'}+\d_{[a|}^{[c}L_{b]}{}^e{}_{e'}W_e{}^{d]e'}+\\ 
                 &+\d_{[a|}^{[c}W_{b]}{}^e{}_{e'}L_e{}^{d]e'})+\\
                   &+\frac{2}{(k-1)(k-2)}\d_{[a}^c\d_{b]}^d(W^2-L^2),\\
	\tilde{C}'_{a'b'}{}^{c'd'}=&C'_{a'b'}{}^{c'd'}-\frac{4}{k'-2}\d^{[c'}_{[a'}{C'}^{d']}_{b']}
                +\frac{2}{(k'-1)(k'-2)}C'-\\
                  &- 2(L'_{[a'|}{}^{c'}{}_{e}L'_{b']}{}^{d'e}+2L'_{[a'|}{}^{[c'}{}_{e}W'_{b']}{}^{d']e})-\\
                  & -\frac{4}{k'-2}(\d_{[a'|}^{[c'}L'_{b']}{}^{e'}{}_{e}L'_{e'}{}^{d']e}+
                   \d_{[a'|}^{[c'}W'_{b']}{}^{e'}{}_{e}W'_{e'}{}^{d']e}+\d_{[a'|}^{[c'}L'_{b']}{}^{e'}{}_{e}W'_{e'}{}^{d']e}+ \\ 
                 &+\d_{[a'|}^{[c'}W'_{b']}{}^{e'}{}_{e}L'_{e'}{}^{d']e})+\\
                   &+\frac{2}{(k'-1)(k'-2)}\d_{[a'}^{c'}\d_{b']}^{d'}({W'}^2-{L'}^2)\\
}
	where 
	\eqal{3}{
               	C_{ab}{}^{cd}:=&\tC_{ab}{}^{cd},& C_{a}^{b}:=&C_{ac}{}^{bc},& C:=&C_{ab}{}^{ab}\\
        	C'{}_{a'b'}{}^{c'd'}:=&\tC_{a'b'}{}^{c'd'},&\qquad C'{}_{a'}^{b'}:=&C'_{a'c'}{}^{b'c'},&\qquad C':=&C'_{a'b'}{}^{a'b'}.
       	}
}
From this proposition it follows that if the semi-basic part of the conformal tensor is zero and the distribution is integrable then the distribution
possess local conformal flatness if and only if its conformation tensor vanishes. Of course this is only true in the case where $k>3$. This generalizes
Carter's result to the case of almost product manifold.
%
%
\section{Physical applications}

There are lots of physical applications involving almost product manifolds. Because principal bundles can be regarded as a almost product manifold
with the $(GF,GD)$ structure ordinary gauge theory can be found in its utmost geometrical form. In Kaluza--Klein theory the internal space
need not be a group manifold but could instead be a homogenous space with the proper gauge group as its isometry group. Here will be given
an example of the recovered Kaluza--Klein theory from the almost product structure taken first in the  most general case where no restriction of
the fiber is made.
\exam{Kaluza-Klein}{{\bf Kaluza-Klein theory}\\
Here will be seen how, in the case of a $(GF,GD)$ structure, the Vidal connection will reduce to the gauge covariant derivative.
The Einstein--Hilbert action will reduce to the inner curvatures plus the gauge field term as in \cite{DuNiPo85}. First note the spliting of the action
in the $(GF,GD)$ case,
\eq{
       \int d^mx\sqrt{\tg}\tR=\int d^kxd^{k'}\!\!y\sqrt{\ig}\sqrt{\nlg}(\Tilde{\Tilde{R}}+\Tilde{\Tilde{R}}'+L^2).
}
Note that the primed distribution is chosen to be integrable. Following \cite{DuNiPo85} the vielbeins can locally be parametrized as
\eqal{2}{
          \tE_a=&E_a+A_a^iK_i,\quad&\tE_{a'}&=E_{a'}\\
          \tE^a=&E^a,     &\tE^{a'}=&E^{a'}-E^aA_{a}^iK_i^{a'}
}
where $K_i=K_i{}^{a'}(y)E_{a'}(y)$ are the Killing vectors of the integrable internal manifold. These satisfy an algebra
\eq{
         [K_i,K_j]=f_{ij}{}^kK_k.
}
where the structure constants $f_{ij}{}^{k}$ of the isometry group and does not depend on $y$. The gauge fieldstrength 
$F^i=dA^i+{1\fr 2}f_{jk}{}^iA^j\wedge A^k$ is most easily found \cite{Holm98}
\eq{
	F(X,Y)=\nco [X,Y]=2L(X,Y)
}
where $X,Y$ are unprimed vector fields which implies that the $L^2$ term in the action reads ${1\fr 4}F^2$ which is the ordinary action term in 
gauge theory. Now the Vidal connection of the gauge field can be written
\eq{
	(\rdel_X F)(Y,Z)=(\tdel^{\esD}_XF)(Y,Z)+F^i(Y,Z)\nco [X,K_i],
}
written in component form, the relations look like,
\eq{
	(\rdel_aF)_{bc}^i=(\tdel^{\esD}_aF)_{bc}^i+f_{jk}^iA_a^jF_{bc}^k
}
which is precisely the gauge covariant derivative. Further the identity from proposition \ref{twistbianchi} $(\rdel_{[a}L)_{bc]}{}^{d'}=0$ reduces
to the Bianchi identity of the gauge field
\eq{
	(\rdel_{[a}F)_{bc]}^i=0
}
the $\tR_{ab'}$ term from proposition \ref{Riccividal} reduces to
\eq{
	\tR_{ab'}={1\fr 2}(\rdel_cF)_{a}^i{}^{c}K_{ib'}
}
which from the Einstein's equations point of view reduces to the equations of motion for the gauge field, i.e. $(\rdel_cF)_{a}^i{}^{c}=0$.
So it is clear that gauge theory and Kaluza-Klein theory is contained in the almost product manifold description. In the general case however
the Killing vectors could be exchanged to ordinary vielbeins, the structure coefficients need not be constant and the fiber no longer a
group space or homogenous space, the almost product structure procedure would still be valid. In the case of Kaluza-Klein theory containing
the dilaton field it is easy to see that it is contained in the mean curvature, $\k$, see \cite{Holm98}.
}
Another more traditional example is the decomposition of the four dimensional curvature scalar into three space in hamiltonian formulation
of ordinary Einstein gravity. From proposition \ref{Ricciadapted} the decomposition of the curvature scalar is immediately found to be $\tR=\aR-\k^2+K^2$.
From a foliation point of view it is clear that a space with non-degenerate metric of Minkowskian signature must have vanishing euler number
which is also the condition for a space to have a codimension one foliation \cite{To88,Rov98}. This is why the $L^2$ term can be set to zero.
%
%
\section{Conclusions and outlook}

The theory of almost product manifolds is seen to overlap with a lot of physical applications. The main areas is of course geometrical phyics
such as gravity, Kaluza-Klein theory and ordinary gauge theory. Here was seen for instance that in gauge- or Kaluza-Klein theory
the Vidal connection reduces to the ordinary gauge covariant derivative and the second curvature identity of the Vidal curvature gives the Bianchi
identity of the gauge field. The relations found propositions \ref{Ricciadapted} and \ref{Riccividal} could perhaps be used to find new solutions
to the equations of motions of various supergravity theories. Here the the Ricci tensors are given in the most general case why all kind of brane solutions
must fit in this scheme. From these relations it should also be clear that a black hole solution in ordinary space-time carrying a gauge field charge
should correspond to a black hole solution in the total space with rotational parameters corresponding to the gauge charges. The reason for this is of
course the identification of the twisting tensor as the gauge fieldstrength. In super gravity theories the black hole solutions carry charges from
anti-symmetric tensor fields and will be $p$-brane solutions. These can again have rotational parameters in the transverse directions
which correspond to gauge field charges of the isometry group of the transverse space \cite{CvDu99,Du98} which by analysis in \cite{Holm98} would correspond
to a non-integrability of the brane itself in this context. 
 
Another interesting investigation would be to see how the Clifford algebra splits in a almost product manifold.
In a appearing paper \cite{Mig} will be shown how flat super space looks in the almost product structure picture.
\vskip1cm
\ul{Acknowledgments}:
The authors would like to thank Martin Cederwall and Robert Marnelius for discussions.
\appendix
%
%
\section{Tic--Tac--Toe notation}

The Tic-Tac-Toe notation can most easily be described by working through the first non-trivial Young tableau, which is
the (2,1) one. 
\eq{
	\yng(2,1)=\young(oo,o) \oplus \young(ox,o) \oplus \young(oo,x) \oplus \young(xx,o) \oplus \young(xo,x) \oplus \young(xx,x)
} 
Above the Tic-Tac-Toe notation was used where the $o$'s labels unprimed degrees of freedom and the $x$'s primed ones. The Tic-Tac-Toe 
tableaux works exactly as an ordinary Young-tableau. Since the primed and the unprimed directions 
does not talk with each other when it comes to symmetries, the dimension of a Tic-Tac-Toe tableau equals
the product of the dimensions of the pure primed and unprimed sub-tableaux respectively. Their respective dimensions read
\eqal{1}{
	\frac{m(m^2-1)}{3}=&\frac{k(k^2-1)}{3}+\frac{k(k-1)k'}{2}+\frac{k(k+1)k'}{2}+\\
			&+\frac{k'(k'+1)k}{2}+\frac{k'(k'-1)k}{2}+\frac{k'({k'}^2-1)}{3}
}
where $k'=m-k$.
The decomposition of an arbitrary Young tableau can be done in a similar fashion. 

\def\href#1#2{#2}
\begingroup\raggedright\endgroup

\end{document}